\documentclass[twocolumn]{aastex631}{}
\usepackage{amsmath}
\usepackage{subfigure}
\usepackage{float}

\newcommand{\sgra}{Sgr~A$^*$}

\begin{document}

\title{Advancing Black Hole Imaging with Space-Based Interferometry}

\author[0000-0003-0387-6617]{Yassine BEN ZINEB}
\affiliation{Georgia Institute of Technology, School of Physics, 837 State St NW, Atlanta, GA 30332, USA}

\author[0000-0003-4413-1523]{Feryal OZEL}
\affiliation{Georgia Institute of Technology, School of Physics, 837 State St NW, Atlanta, GA 30332, USA}

\author[0000-0003-1035-3240]{Dimitrios PSALTIS}
\affiliation{Georgia Institute of Technology, School of Physics, 837 State St NW, Atlanta, GA 30332, USA}

\begin{abstract}
Horizon-scale imaging with the Event Horizon Telescope (EHT) has provided transformative insights into supermassive black holes but its resolution and scope are limited by ground-based constraints such as the size of the Earth, its relatively slow rotation, and atmospheric delays. Space-based very long baseline interferometry (VLBI) offers the capability for studying a larger and more diverse sample of black holes. We identify a number of nearby supermassive black holes as prime candidates for horizon-scale imaging at millimeter wavelengths, and use source characteristics such as angular size, sky distribution, and variability timescales to shape the design of a space-based array. We identify specific metrics that serve as key predictors of image fidelity and scientific potential, providing a quantitative basis for optimizing mission design parameters. Our analysis demonstrates that the optimal configuration requires two space-based elements in high Earth orbits (HEO) that are not coplanar and are apparently counter-rotating. Our results delineate the key requirements for a space-based VLBI mission, enabling detailed studies of black hole shadows, plasma dynamics, and jet formation, advancing black hole astrophysics beyond the current capabilities of the EHT.
\end{abstract}


\section{Introduction} \label{sec:intro}

Horizon scale imaging with the Event Horizon Telescope (EHT) opened a new path for studying the gravitational fields and accretion physics of two supermassive black holes, M87 \citep{EHT2019_I} and \sgra\ \citep{EHT2022_I}. The ring-like structures in the images enabled new tests General Relativistic predictions \citep{Psaltis2020,EHT2022_VI}, while the early polarization, variability, and brightness asymmetry results provided new probes of plasma conditions and magnetic field structures \citep{EHT2019_V,EHT2021_VII,EHT2022_V,EHT2024_VII,Satapathy2022,Faggert2024}.

Imaging with very long baseline interferometry from the ground has reached its limit of maximum possible resolution with the inclusion of stations from Greenland to the South Pole in the N-S direction and from Hawaii to the French Alps in the E-W direction. This limits horizon-scale imaging to the two sources that have already been observed \citep{EHT2022_II}. Moreover, ground-based observing has other natural challenges, such as the loss of phase coherence at short wavelengths due to the atmosphere \citep{EHT2019_III}, the presence of narrow atmospheric transmission windows for observations, and the reliance on the rotation of the Earth to enable (limited) coverage of the Fourier space. Going beyond these constraints and obtaining horizon-scale images for a large sample of supermassive black holes require the large baselines that only space arrays can provide. 

\begin{deluxetable*}{lcccccccl}[t]
\tablecolumns{9}
\tablecaption{Black Hole Targets\label{tab:candidates}}
\tablehead{\colhead{Name}&\colhead{Flux}&\colhead{Mass}&\colhead{Distance}&\colhead{Ring}&\colhead{Variability}&\colhead{Right Ascension}&\colhead{Declination}&Refs. \\[-0.3cm]
&&&&\colhead{Diameter}&\colhead{Timescale}&&& \\[-0.3cm]
& \colhead{(Jy)} & \colhead{($ 10^8~M_\odot$)} & \colhead{(Mpc)} & \colhead{($\mu$as)} & \colhead{(days)}&{} &{}&{}}
\startdata
\sgra\ & $2.4$ & $0.043$ & $0.0083$ & $53.14$ & 0.7 & $17^h \, 45' \, 40.0383''  $ & $-29^\circ \, 0' \, 28.069''$ & 1 \\[0.1cm]
M87 & $0.5$ & $62$ & $16.8$ & $37.85$ & $35.5$ & $12^h \, 30' \, 49.4233''  $ & $12^\circ \, 23' \, 28.043''$ & 2 \\[0.1cm]
M60$^\dagger$ & $1.7 \times 10^{-3}$ & $45^{+10}_{-10}$ & $17.38$ & $26.56$ & $29.8$ & $12^h \, 43' \, 39.9948''  $ & $11^\circ \, 33' \, 9.402''$ & 3, 4\\[0.1cm]
IC 1459 & $0.217$ & $26^{+11}_{-11}$ & $26.18$ & $10.19$ & $22.02$ & $22^h \, 57' \, 10.6068''  $ & $-36^\circ \, 27' \, 44''$ & 5, 6\\[0.1cm]
NGC 3115$^\dagger$ & $\leq 2.07 \times 10^{-4}$ & $9.6^{+0.9}_{-1.9}$ & $10.3$ & $9.56$ & $12.7$ & $10^h \, 5' \, 13.9780''  $ & $-7^\circ \, 43' \, 6.891''$ & 5, 7\\[0.1cm]
NGC 4594 & $0.198$ & $6.6^{+0.4}_{-0.4}$ & $9.55$ & $7.09$ & $10.4$& $12^h \, 39' \, 59.4318''  $ & $-11^\circ \, 37' \, 22.996''$ & 8, 9\\[0.1cm]
NGC 3998 & $0.13$ & $8.1^{+2}_{-1.9}$ & $14.2$ & $5.85$ & $11.6$ & $11^h \, 57' \, 56.133''  $ & $55^\circ \, 27' \, 12.922''$ & 5, 10\\[0.1cm]
NGC 2663 & $0.084$ & $16$ & $28.5$ & $5.76$ & $16.9$ & $8^h \, 45' \, 8.144''$ & $-33^\circ \, 47' \, 41.064''$ & 11, 12 \\[0.1cm]
NGC 4261 & $0.2$ & $16.7^{+4.9}_{-3.4}$ & $32.4$ & $5.29$ & $17.3$ & $12^h \, 19' \, 23.2204''  $ & $5^\circ \, 49' \, 30.775''$ & 5, 13\\[0.1cm]
M84 & $0.133$ & $8.5^{+0.9}_{-0.8}$ & $17$ & $5.13$ & $11.9$ & $12^h \, 25' \, 3.7433''  $ & $12^\circ \, 53' \, 13.139''$ & 5, 14\\[0.1cm]
NGC 3894 & $0.058$ & $20$ & $48.2$ & $4.26$ & $19.1$ & $11^h \, 48' \, 50.3582''  $ & $59^\circ \, 24' \, 56.382''$ & 15, 16\\[0.1cm]
3C 317 & $0.034$ & $46^{+3}_{-3}$& $122$ & $3.85$ & $30.1$ & $15^h \, 16' \, 44.507''$ & $7^\circ \, 01' \, 18.078 ''$ & 5, 17 \\[0.1cm]
NGC 4552 & $0.027$ & $4.9^{+0.7}_{-0.4}$ & $16.3$ & $3.08$ & $8.8$ & $12^h \, 35' \, 39.8073''  $ & $12^\circ \, 33' \, 22.829''$ & 18, 19\\[0.1cm]
NGC 315 & $0.182$ & $20.8^{+3.3}_{-1.4}$ & $70$ & $3.05$ & $19.5$ & $0^h \, 57' \, 48.8833''  $ & $30^\circ \, 21' \, 8.812''$ & 13\\[0.1cm]
NGC 1218 & $0.11$ & $31.6$ & $116$ & $2.8$ & $24.5$ & $3^h \, 8' \, 26.2238''  $ & $4^\circ \, 6' \, 39.3''$ & 20, 21\\[0.1cm]
IC 4296 & $0.212$ & $13.4^{+2.1}_{-1.9}$ & $50.8$ & $2.71$ & $15.3$ & $13^h \, 36' \, 39.0330''  $ & $-33^\circ \, 57' \, 57.073''$ & 22, 23 \\[0.1cm]
NGC 1399 & $0.038$ & $5.1^{+0.7}_{-0.7}$ & $21.1$ & $2.48$ & $9$ & $3^h \, 38' \, 29.0200''  $ & $-35^\circ \, 27' \, 0.700''$ & 24, 25 \\[0.1cm]
Cen A & $5.98$ & $0.55^{+0.3}_{-0.3}$ & $3.42$ & $1.65$ & $2.6$ & $13^h \, 25' \, 27.6150''  $ & $-43^\circ \, 1' \, 8.806''$ & 26, 27\\[0.1cm]
\enddata
\tablerefs{1. \citet{GRAVITY2023_flare} 2. \citet{Akiyama_2019_bis} 3. \citet{Tully_2016} 4. \citet{Shen_2010} 5. \citet{Tully_2013} 6. \citet{Cappellari_2002} 7. \citet{Gültekin_2009} 8. \citet{Jardel2011} 9. \citet{McQuinn_2017}  10. \citet{Walsh2012} 11. \citet{ngEHT} 12. \citet{Velovic_2012} 13. \citet{Boizelle2021} 14. \citet{Walsh_2010} 15. \citet{2007...465...71T} 16. \citet{1999ApJ...519..117C} 17. \citet{Mezcua_2017} 18. \citet{2010ApJ...717..603V} 19. \citet{Saglia_2016} 20. \citet{Crook2007} 21. \citet{1986MNRAS.219..545S} 22. \citet{Jensen_2003} 23. \citet{Bontà_2009} 24. \citet{Tonry_2001} 25. \citet{Gebhardt_2007} 26. \citet{Cappellari_2009} 27. \citet{Ferrarese_2007}}
\tablecomments{The two sources marked with a dagger are too faint to be detected during an observation campaign and we chose to keep them out of the rest of our analysis}
\end{deluxetable*}

High-resolution images can provide a wealth of information about accretion physics, gravity, and the astrophysics of black holes. Image properties are dictated by the structure of organized magnetic fields around black holes \citep{EHT2021_VIII,Narayan2021,EHT2024_VIII}, the plasma stresses that determine densities and velocities \citep{EHT2019_V,Medeiros2022,EHT2022_V}, the presence of small- and large-scale turbulence \citep{Satapathy2022,Wielgus2022}, and the formation and loading of powerful jets \citep{Lu2023}. Furthermore, the strong-field metrics of black holes are imprinted on the size and asymmetry of their shadows \citep{EHT2019_VI,Psaltis2020,EHT2022_VI}. Finally, the same image characteristics lead to the most precise measurements of black hole masses beyond the Milky Way and a direct probe of the alignment of black hole spins with large scale galactic structures \citep{Johannsen2012,Psaltis2015,EHT2019_VI}. 

\begin{figure*}[t]
    \centering
    \includegraphics[width=\textwidth]{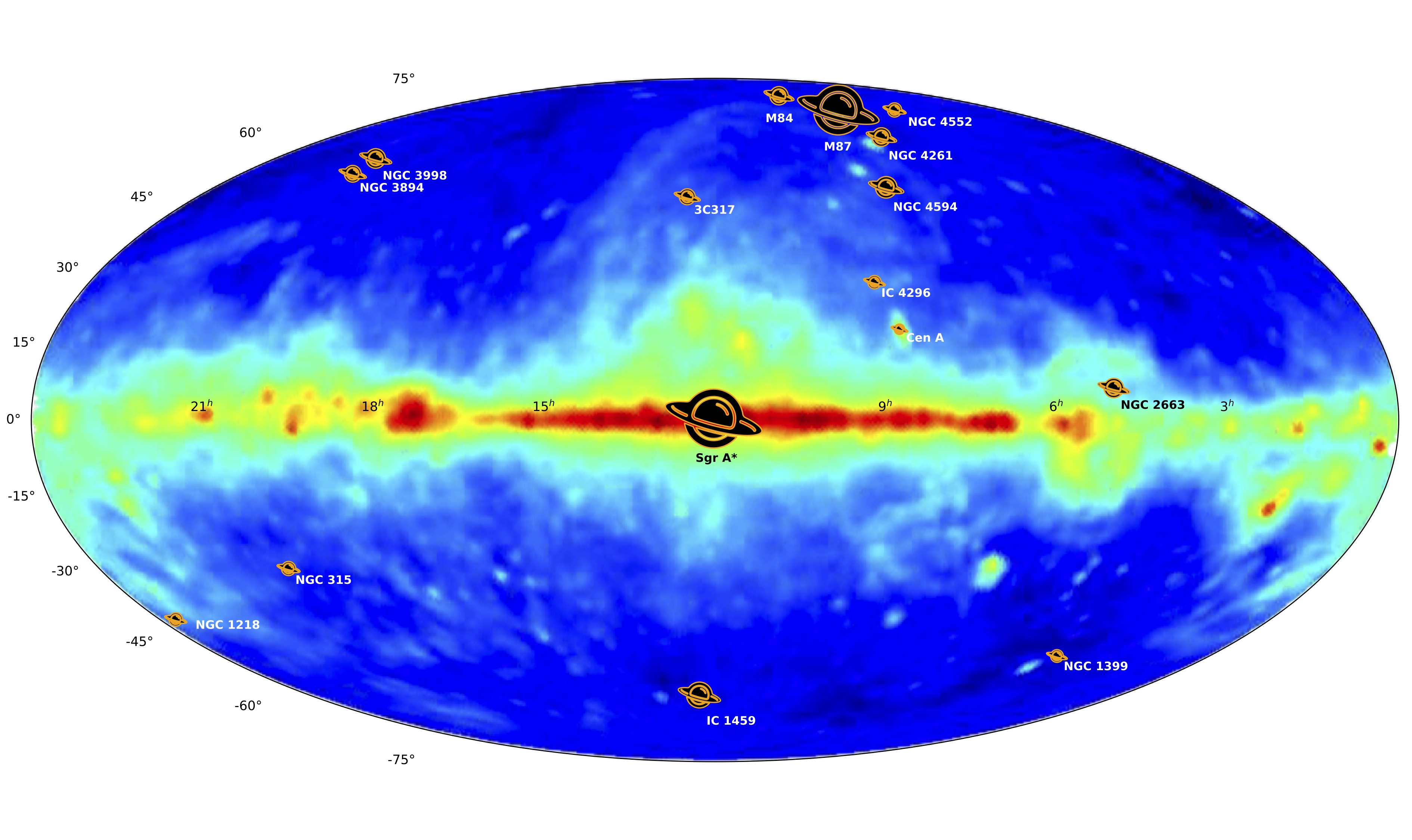}
    \caption{Positions of 16 black hole targets in the sky with the largest known angular sizes and millimeter fluxes. For clarity, M84 and NGC4552 have been slightly displaced to avoid overlap with M87. The marker size of each black hole is linearly scaled based on the mass-to-distance ratio. The sources span a high diversity of directions in the sky and require an array configuration that is not uni-directional to image them. The background image is a 24~GHz all sky map from \citet{Zhang_2009}}
    \label{fig:sourceposition}
\end{figure*}

The current EHT observations of the two primary targets revealed a number of surprising results in these regards. The nearly symmetric rings observed from both sources require either a highly coincidental line-up between the spin axis of each black hole and our line of sight or the presence of strong stresses that slow down plasma motion near the horizons \citep{Faggert2024}. The unexpected lack of high-amplitude variability at the dynamical timescale in both sources requires a new understanding of plasma turbulence and its effect on the emission of radiation \citep{Satapathy2022,EHT2022_V}. Just as importantly, the similarity between the images of a jetted (M87) and non-jetted (\sgra) source raises questions about the physics of horizon-scale loading of jets \citep{Lu2023}. Pinning down the prevalence and universality of these phenomena requires the detailed horizon-scale images of a large sample of black holes. 

A number of early concepts have been explored to extend the imaging capabilities beyond those of the EHT. Some studies focused on adding a telescope in orbit to augment the existing Earth-based array (see, e.g., \citealt{Andrianov2021}), while others utilized only space-based elements in varying numbers and orbital altitudes \citep{Trippe2023,Roelofs2019,MartinNeira2022,Kudriashov2021}. To increase resolution, especially at relatively low-Earth altitudes that do not increase the baseline length appreciably with respect to the EHT, some of these studies rely purely on increasing the observing frequency \citep{Trippe2023}, while others reach large altitudes (e.g., halo orbits around the L2; see \citealt{Andrianov2021}). These studies have focused almost exclusively on increasing resolution for the primary EHT targets and did not systematically explore the requirements necessary to comprehensively image a broad sample of targets. 

In this paper, we first identify a broad range of targets with characteristics that are well suited for high resolution imaging with a space-based interferometer and will enable the science goals outlined above. We then present a framework to map orbital characteristics to imaging capabilities for black hole sources. Focusing on the fidelity of reconstructed images, we apply this framework to the sample of targets we have identified and carry out a high-level trade study. We use this to delineate the basic characteristics of a space mission that enables a range of science goals. 

\section{Black Hole Target Selection and Modeling}

We focus on the properties of nearby supermassive black holes to find viable candidates for horizon-scale imaging at millimeter wavelengths. Several source characteristics will determine the general design specifications of an interferometric array: the angular size of the horizons will set requirements for the baseline lengths; the distribution of the sources in the sky will inform the orientations of the orbits, and the timescale at which image characteristics vary near the horizon will determine the maximum time available for aperture synthesis interferometry. 

\subsection{Candidate Selection and Characteristics}

\citet{Johannsen2012} made the first selection of black hole targets that are optimal for horizon-scale imaging with a space-based interferometer. This was further extended by the Event Horizon and Environs (ETHER) survey \citep{galaxies11010015}. We use these compilations to assess sources on their angular size, millimeter flux, and time variability characteristics. 

Our first selection metric is the angular diameter of the black hole shadow, which we approximate using the Schwarzschild metric expression 
\begin{equation}
    \theta_{\rm ring} = 2 \sqrt{27}\dfrac{GM_{BH}}{c^2 D_{BH}}\;.
\end{equation}
Here $D_{\rm BH}$ is the distance to the black hole, $M_{\rm BH}$ is its mass, and $G$ and $c$ are the gravitational constant and speed of light, respectively. This quantity measures the required resolution that an array needs to achieve to be able to resolve the shadow of each black hole. 

We identify in Table~\ref{tab:candidates} the known black hole sources with the largest angular sizes and sufficiently high millimeter fluxes that make them good candidates for horizon-scale imaging by the first generation of space-based interferometers. To visualize the distribution of these black hole targets in the sky, we also show in Figure~\ref{fig:sourceposition} a Mollweide projection depicting their galactic coordinates. It is evident from this figure that the candidate black holes span a very broad range of directions in the sky. As we will discuss in the later sections, this places important requirements on the design of a space-based array in order to provide good $u-v$ coverage for sources in different orientations in the sky. A positive outcome of this requirement is that a configuration that is able to image the known sources would also be versatile and adaptable to observe new candidates throughout the sky as the opportunities arise. We also note that the region around M87 is  populated with five potential targets, making it a high priority zone for array optimization. 

Finally, a last attribute to consider is the characteristic variability timescale that will constitute an upper limit on the allowed observation time for aperture synthesis to generate a reliable image reconstruction. Early expectations were that this limiting timescale is comparable to the dynamical timescale near the innermost stable circular orbit~\citep{Medeiros2017,Medeiros2018,Roelofs2017}. However, EHT observations of both M87 \citep{Satapathy2022} and \sgra\ \citep{EHT2022_II,Wielgus2022} showed a remarkable lack of variability in the image structure at these timescales.

In light of these results, we instead assume that the morphology of the horizon-scale image can change on the same timescale over which the source flux varies. We calculate this timescale using the empirical relationship between the flux variability and black hole mass found by \citet{2023ApJ95193C} 
\begin{equation}
    \tau \approx 3.67^{+1.07}_{-1.56} \text{days} \left(\displaystyle{\frac{M_{BH}}{10^8 M_\odot}} \right)^{0.55^{+0.09}_{-0.09}}\;.
    \label{eq:Variability}
\end{equation}
This timescale is of order $\sim 10-20$~days for most of the candidates shown in Table~\ref{tab:candidates}. We evaluate the coverage of the $u-v$ plane within this timescale for each source.

\subsection{Requirements for Image Fidelity}

In addition to the largest baseline length that sets the maximum image resolution, imaging requires a reasonable coverage of the $u-v$ space across both baseline lengths and azimuthal orientations. Because the primary purpose of the space interferometer will be capturing the ring-like horizon-scale images of black holes, we will use a simple geometric model to evaluate the requirements that facilitate the highest fidelity reconstructions of such images. 

To this end, we use the geometric crescent model of \citet{10.1093mnrasstt1068}, which describes a crescent as a subtraction of two unit disks and allows for a brightness asymmetry by displacing their centers. The Fourier transform of such an image is given by 
\begin{eqnarray}
    \dfrac{V_c}{V_0} &=&  
     \frac{2 \pi}{k (R_p^2 - R_n^2)} \left[R_p J_1(k R_p) \right.\nonumber\\
     &&\qquad\qquad\left.- e^{-2 \pi i (a u+b v)/ \lambda} R_n J_1 (k R_n)\right]\;,
    \label{eq:crescent}
\end{eqnarray}
where $R_p$ and $R_n$ are radii of the outer and inner disks, respectively, $k = 2 \pi \sqrt{u^2 + v^2}/\lambda$, $J_1$ is the Bessel function of the first kind, and $a$ and $b$ are the displacements in the $u$ and $v$ directions. The parameters of this model can be recast into a set that provides a more intuitive description of the crescent properties: the angular diameter ($\theta_{\rm ring}$), the fractional thickness ($\psi$), the degree of symmetry ($\tau$), and the orientation ($\phi$),
\begin{eqnarray}
     \theta_{\rm ring} &\equiv& R_p+R_n,\\
     \psi &\equiv& 1- \dfrac{R_n}{R_p},\\ 
     \tau &\equiv& 1 - \dfrac{\sqrt{a^2 + b^2}}{R_p -R_n},\\
    \phi &\equiv& \tan^{-1} \dfrac{b}{a}\;.
\end{eqnarray}
We then write Equation~\eqref{eq:crescent} as a function of these more representative variables
\begin{equation}
\begin{array}{ll}
    \dfrac{V_c}{V_0} &=  \dfrac{4 \pi}{\theta_{\rm ring} k \psi (2- \psi)} \left\{J_1(k\theta_{\rm ring}/2) - \right. \\
     & (1-\psi) \exp\left[- \dfrac{i \pi \theta_{\rm ring}\psi (1-\tau) (u\cos \phi + v\sin \phi)}{\lambda} \right] \\
     & \left. J_1[k(1-\psi)\theta_{\rm ring}/2] \right\}.
\end{array}
\end{equation}
In the $\psi\rightarrow 0$ limit, this function describes a ring of infinitesimal thickness, i.e., \begin{equation}
    \dfrac{V_{r}}{V_0} = J_0 (k \theta_{\rm ring}/2)
    \label{eq:infinitesimal_ring}\;.
\end{equation}

\begin{figure}[t]
    \centering
    \includegraphics[width=0.98\linewidth]{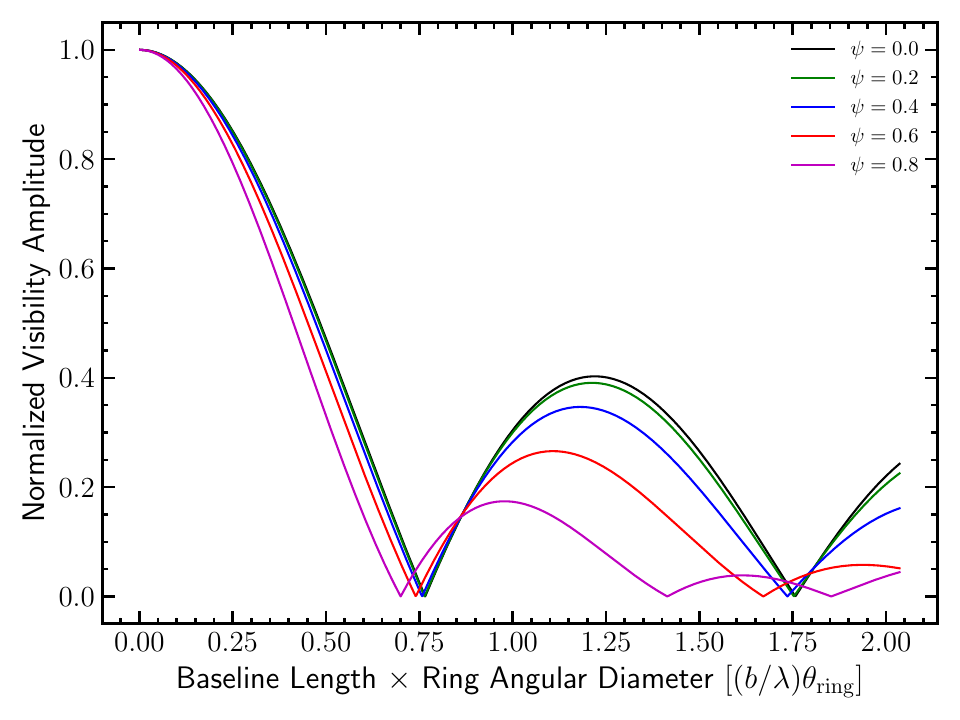}
    \caption{Normalized visibility amplitude for rings of various fractional thicknesses  ($\psi$) plotted against baseline length multiplied by the angular diameter of the ring. The baseline lengths corresponding to the first null, first peak and second null of the visibility correspond to regions that provide most information on the functional shape and are, therefore, of high importance in the u-v plane.}
    \label{fig:visibility}
\end{figure}

We show in Figure~\ref{fig:visibility} the visibility amplitude as a function of baseline length for crescents of various fractional widths for a fixed source size. It is evident from this figure that the first minimum in the visibility amplitude is determined primarily by the size of the ring, with weak dependence on its thickness, while the location and amplitude of the second maximum and the location of the second minimum is highly sensitive to the relative thickness. 

\begin{figure}[t]
    \centering
\includegraphics[width=0.98\linewidth]{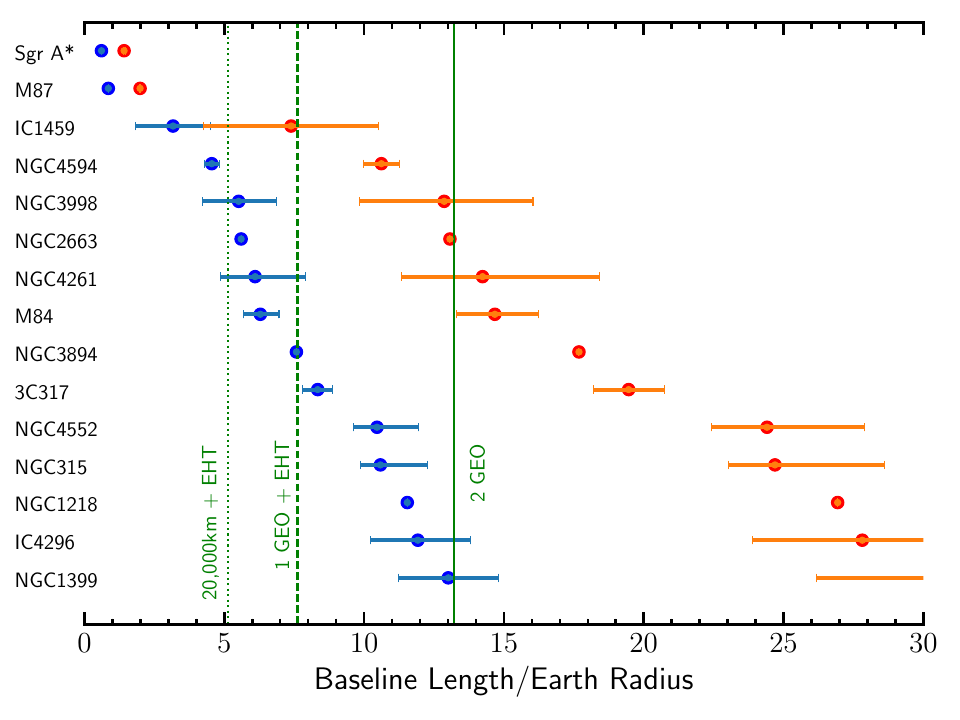}
    \caption{The required baseline lengths to (blue) measure the size of the images and (red) determine the presence of a shadow or other image characteristics at 1.3$\;$mm for the sample of black hole targets shown in Table~\ref{tab:candidates}. The green lines show the maximum baseline lengths that can be achieved by a spacecraft in MEO together with the ground-based EHT, a spacecraft in GEO with the EHT, and two spacecrafts in GEO. Significantly increasing the sample size of black hole images at 1.3$\;$mm requires baseline lengths in excess of $20 \; R_\oplus$, which is possible with an array of at least 2 satellites in HEO.}
    \label{fig:required_baselines}
\end{figure}

These characteristics imply that probing the baseline lengths corresponding to the first minimum 
\begin{equation}
    b_1\simeq \frac{0.75}{\theta_{\rm ring}} \lambda 
    \label{eq:b_1}
\end{equation}
is necessary to measure accurately  the source size. Moreover, probing baseline lengths up to at least the second minimum 
\begin{equation}
    b_2\simeq \frac{1.75}{\theta_{\rm ring}} \lambda
    \label{eq:b_2}
\end{equation}
is required to determine the ring width, which is important for distinguishing an image with and without the presence of a shadow. 

Figure~\ref{fig:required_baselines} shows the baseline lengths corresponding to the first (blue) and second minima (red) for each of the targets included in Table~\ref{tab:candidates}. It also shows the maximum possible baseline of three array configurations: a spacecraft in MEO together with the ground-based EHT, a spacecraft in GEO with the EHT, and two spacecraft in GEO. It is evident from this figure that the first configuration can at most measure the size of one black hole target beyond \sgra\ and M87. Two orbiters in GEO can measure the sizes of all targets but cannot determine the presence of a shadow or any other image characteristic. Increasing the sample size of black hole images by an order of magnitude requires baseline lengths in excess of $20\; R_\oplus$, which is possible with an array of at least 2 satellites in HEO. 

\begin{figure}[t]
    \centering
    \includegraphics[width=0.97\linewidth]{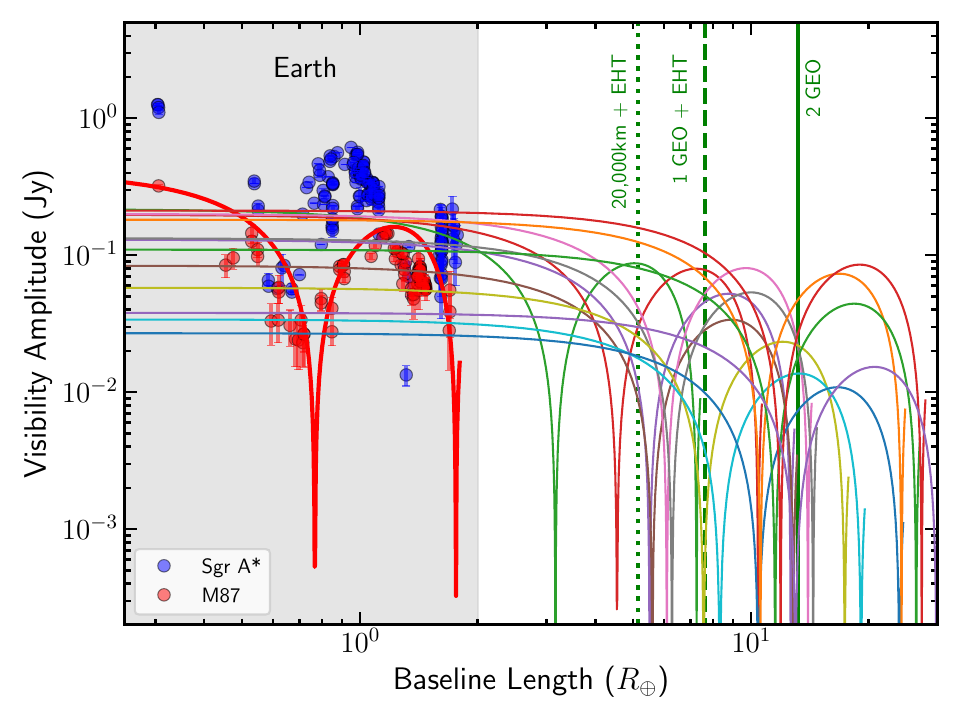}
    \caption{Visibilities of each target listed in Table \ref{tab:candidates} assuming ring-like images of infinitesimal width. \sgra\ and M87 fall in the range of feasibility of the EHT array, as expected. The first minima in the visibility amplitude for most other candidates occur at much larger baseline lengths that necessitate an orbital altitude at least at GEO.}
    \label{fig:max_baselines}
\end{figure}

In the limit of an infinitesimally thin ring, Figure~\ref{fig:max_baselines} shows the expected visibility amplitudes as a function of baseline length for the candidate black hole targets. This figure further highlights the range of baseline lengths that have the highest influence on imaging and are, therefore, required to determine image characteristics. Furthermore, this figure shows that measuring the salient features in the u-v maps of these black hole images requires a correlated flux sensitivity of $\simeq 0.005 \;$Jy.

These considerations allow us to connect the science requirements to orbital radii of satellites and narrow down the range of configurations worth pursuing. However, the requirement of a flexible array configuration to observe black holes throughout the sky imposes further constraints on the relative orientation. Furthermore, choices for the orbits affect the sparsity of $u-v$ coverage, which is critical for high-fidelity image reconstruction. To facilitate these explorations, we will present in the next section the formalism and tools we employ for calculating $u-v$ coverage for arbitrary satellite and array configurations.

\section{Optimizing Space-Array Configurations}

In the preceding sections, we identified an important requirement for obtaining high-fidelity images of black holes using an interferometric array that contains or consists of space elements. This first requirement pertains to probing the salient baselines, i.e., minimally reaching the baseline defined in equation~(\ref{eq:b_2}) for all targets of interest. There is, however, one additional requirement that needs to be taken into account in an array design, which we describe here. 

In order for an image reconstructed from a sparse array to recover even the most broadbrush characteristics of the image without relying on strong priors, the array must provide continuity of coverage both in baseline length and in azimuth such that the largest gaps in the $u-v$ plane are smaller than the correlation length of the image (see \citealt{Psaltis2024a,Psaltis2024b} for details). In particular, for an image with an angular diameter $\theta_{\rm image}$, this leads to a requirement 
\begin{equation}
    b_{\rm gap} \lesssim b_{\rm cor}\simeq \frac{0.45}{\theta_{\rm image}} \lambda\;.
    \label{eq:b_cor}
\end{equation}

\begin{figure*}[ht!]
    \centering
    \includegraphics[width=0.95\textwidth]{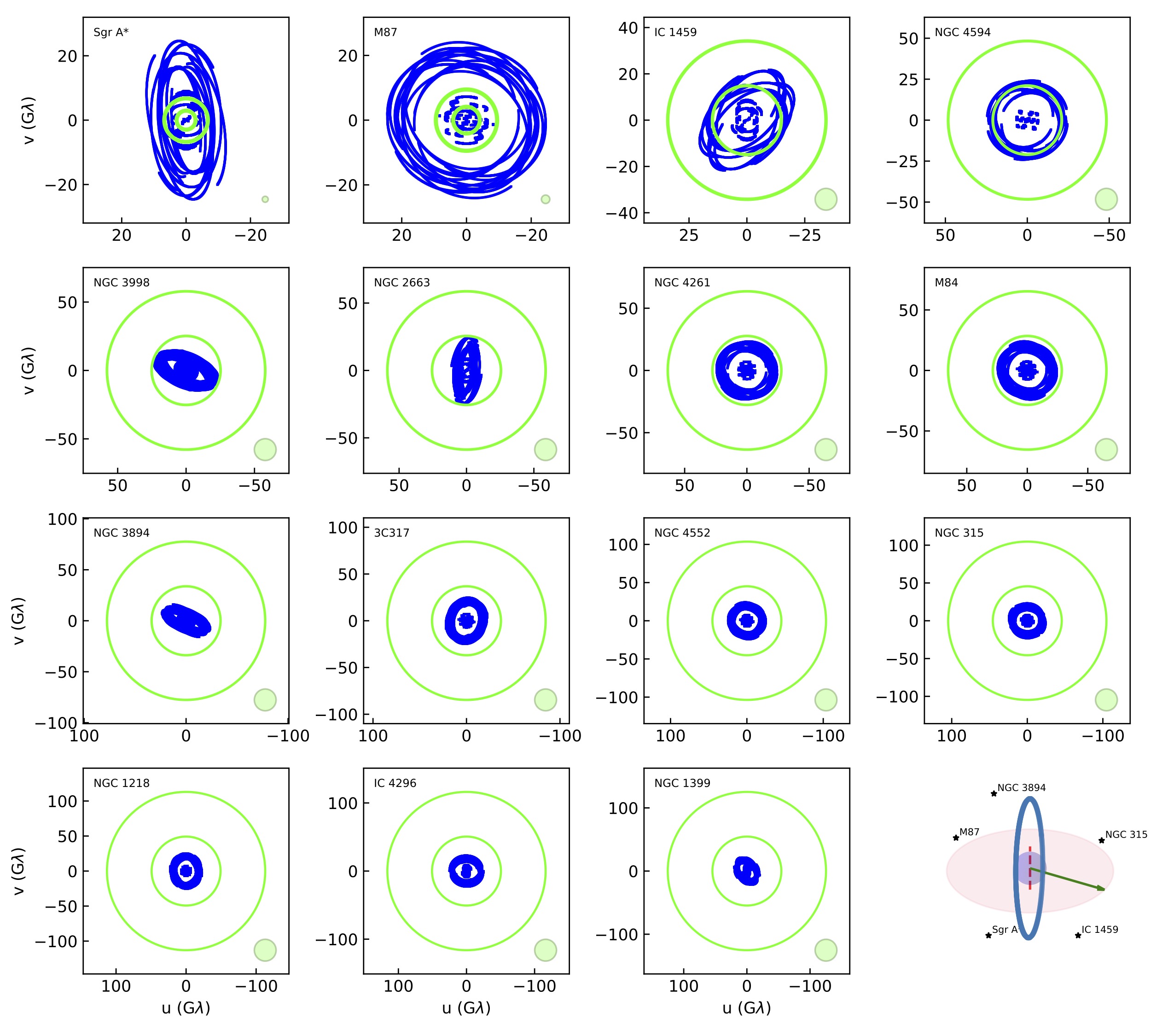}
    \caption{The $(u, v)$ plane coverage for 15 nearby black hole candidates for an array configuration that consists of the EHT ground telescopes and a satellite on a circular orbit at $a=20200$ km with and inclination $i = 85^\circ$ and a longitude of the ascending node $\Omega = 110^\circ$. The targets have been observed for $24$ hours. The green circles indicate the expected locations of the baseline lengths corresponding to the first two minima in the $u-v$ maps of each source. The filled green circle shows the expected correlation length in the $u-v$ map of each source. This configuration does not meet the two criteria required for imaging: reaching the first and second minima (for most sources) and providing sufficiently dense coverage (i.e., avoiding radial or azimuthal gaps larger than a correlation length) for all targets.}
    \label{fig:EHT_MEO}
\end{figure*}

We now use these requirements to assess the ability of a number of hybrid Space-Ground and Space-Space configurations to image the black hole sources identified in Table~\ref{tab:candidates}. We show that this uniquely leads to a space-space configuration with two satellites in counter-rotating and non-coplanar HEO orbits.  

\begin{figure*}[t!]
    \centering
    \includegraphics[width = \textwidth]{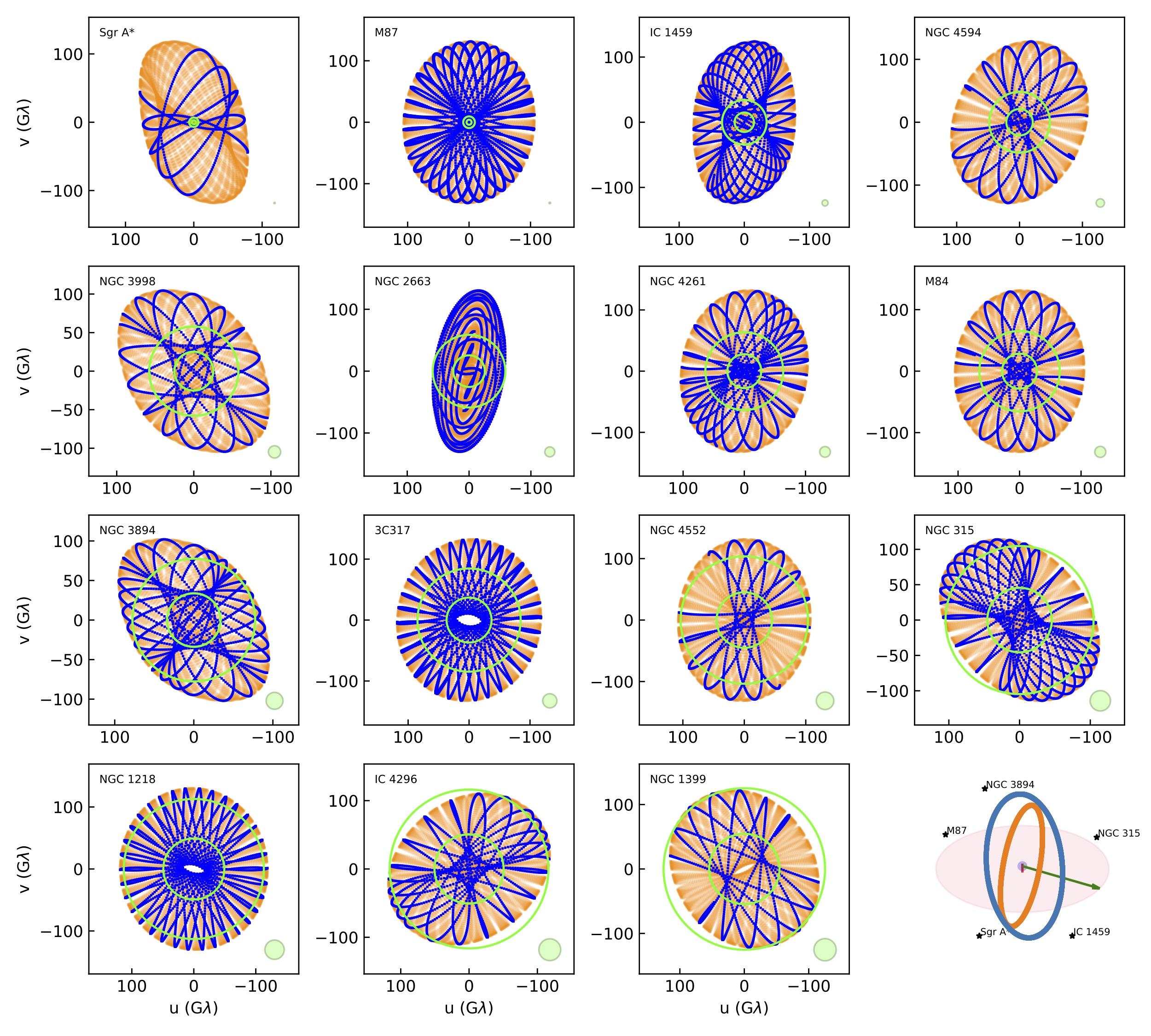}
    \caption{Same as Fig.~\ref{fig:EHT_MEO} but for an array with two satellites on circular orbits of radii $a_1=14.64 R_\oplus$ and $a_2 = 12.5 R_\oplus$. The orbits have been inclined with respect to each other in both angles $i$ and $\Omega$, as shown in the bottom right panel, to optimize sky coverage. We set the difference in the longitudes of the ascending nodes $\Delta \Omega > 90^\circ$ so that the two elements appear to be counter-rotating for most of the targets. The orange curves show the coverage that can be achieved over the course of 6 months. Blue curve trace the coverage that is achieved within the variability timescale of each black hole; for \sgra\ we used $\sim 1\;$week in line with recent EHT observations. This configuration provides dense and nearly isotropic coverage for nearly all targets.}
    \label{fig:2orbiters}
\end{figure*}

\subsection{Space-Ground configurations}

It is already clear from Figure~\ref{fig:required_baselines} that horizon scale images of the candidate black hole targets with quality similar to those of \sgra\ and M87 are not accessible to configurations with orbiting stations that do not reach HEO altitudes. We elucidate this requirement in Figures~\ref{fig:EHT_MEO}, where we show the $u-v$ coverage for an example hybrid array with the existing ground stations plus a station in polar MEO. This inclination is chosen to provide maximum simultaneous coverage with Earth-based stations (see, e.g., \citealt{Johnson2024}). Clearly, the maximum baseline length does not extend beyond the first critical baseline to measure even the size of the image for all but two of the targets beyond M87 and \sgra. 

Figure~\ref{fig:EHT_MEO} serves to illustrate another failure mode in which the configuration does not meet the imaging requirements. Increasing the baseline length by simply moving one array element to a large baseline length introduces significant gaps in the radial coverage of the $u-v$ plane. For the configuration in Figure~\ref{fig:EHT_MEO}, the gaps for M87 are larger than the correlation length of the image. Increasing the maximum baseline further either by increasing the orbital altitude of the space element or by increasing the observing frequency exacerbates these gaps in baseline length. 

Finally, the presence of a single orbiting element necessarily makes the array highly directional. This can be seen in the $u-v$ coverage of \sgra, which has an extremely poor azimuthal coverage in this example. Because the targets are distributed throughout the sky, there is no configuration with a single space based element that overcomes this failure mode. For these reasons, no array configuration that is based on adding a single orbital element to a ground-based array is capable of achieving reasonable imaging and science requirements. We show in the next section that adding a second orbital element makes the ground-based stations obsolete. 
\begin{figure*}[t]
  \centering
  \includegraphics[width=0.329\linewidth]{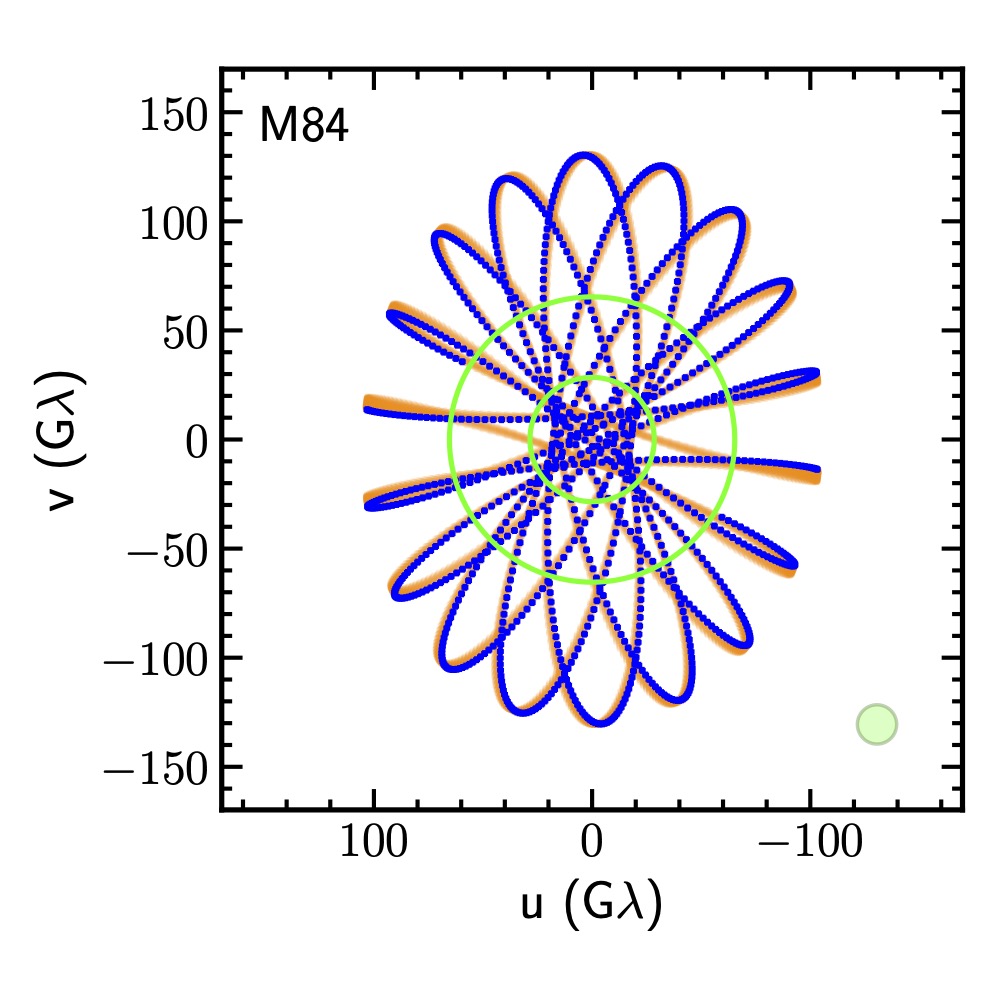}
  \includegraphics[width=0.329\linewidth]{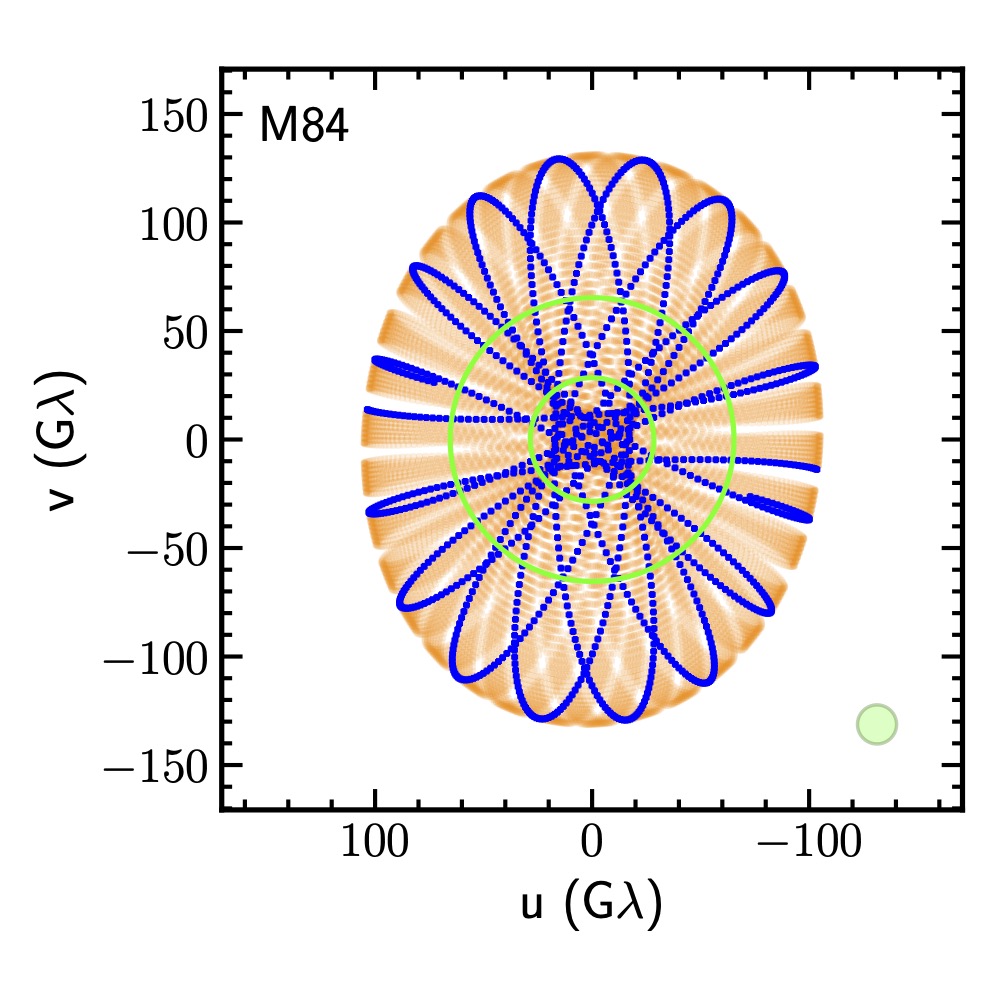}
  \includegraphics[width=0.329\linewidth]{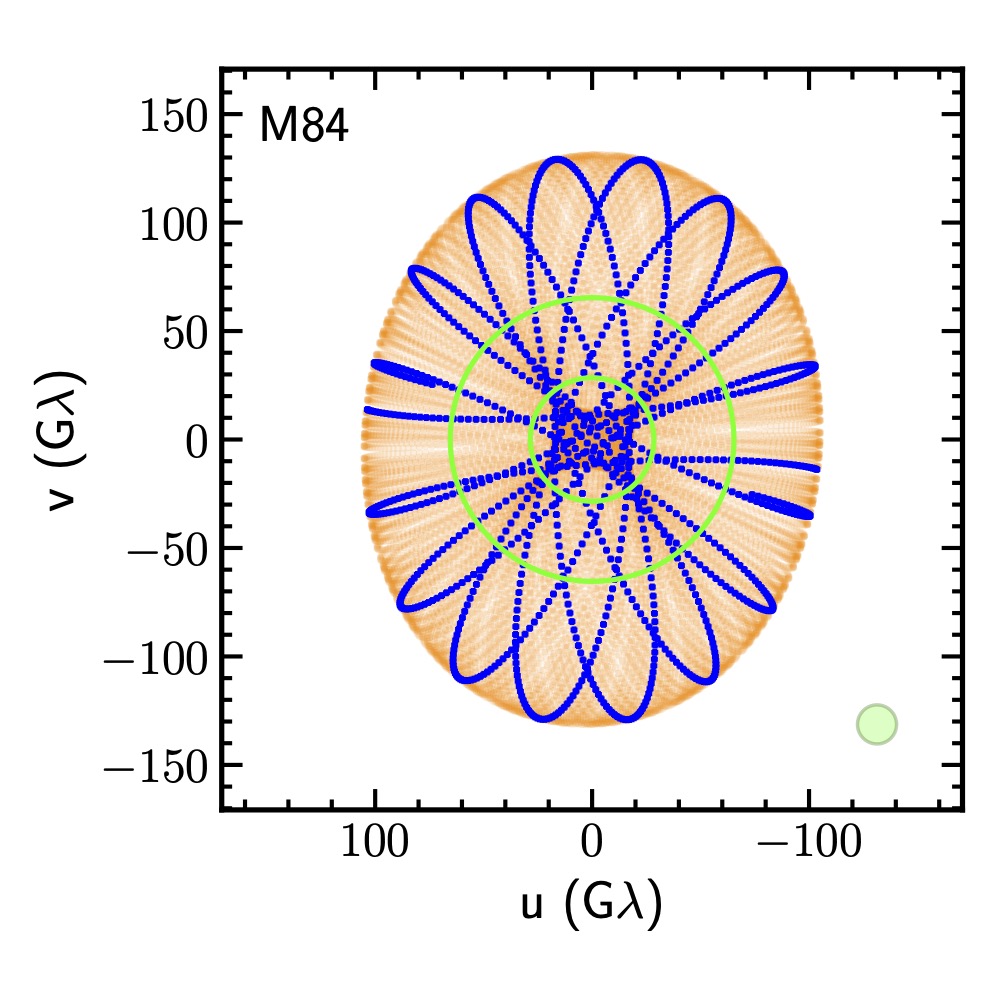}
  \caption{The effect of commensurability between the periods of the orbits on the $u-v$ coverage. The array configuration for the middle panel is the same as in Fig.~\ref{fig:2orbiters} and shows M84 as an example. In the left panel, we set the semimajor axis of the first element to $a_1 = 14.503 \; R_\oplus$, corresponding to a $5/4$ orbital period ratio between the two elements. This reduces the long-term $u-v$ coverage and intensifies the adverse effects of commensurability. In contrast, the right panel shows the case for $a_1 = 14.65$, for which the ratio of periods is irrational.}
  \label{fig:Commensurate}
\end{figure*}

\subsection{Configurations with Two Space Elements} 
\label{sec:TwoOrbiter}

Array configurations with two space elements introduce novel dynamics that significantly fills the $u-v$ space and brings many more sources within reach for imaging. Here, we do not fully optimize the parameters for the orbits but explore the principles that help maximize the science return from such a concept. 

With two space elements, the required baseline lengths shown in Figure~\ref{fig:required_baselines} can be achieved with orbital altitudes that are only half that size. This is because the two elements will appear in diametrically opposed locations during part of their orbits as long as they have different altitudes. As a result, reaching baseline lengths of $\sim 25-30 \; R_\oplus$ requires altitudes of $12-15 \; R_\oplus$. 

The minimum baseline length that can be probed by an array with two space elements is comparable to or smaller than the difference between the two orbital altitudes, depending on the orientation of the source with respect to the two orbits. Placing the two elements in nearby orbits, therefore, minimizes the size of the gaps in baseline lengths for all source orientations, dramatically increasing imaging capability. Furthermore, a difference in orbital altitudes causes a difference in orbital periods that introduces a drift in the separation between the two elements. When this difference is small and the two elements are in counter-rotating orbits, the coverage of the $u-v$ plane becomes dense and fast. Two elements in apparently counter-rotating orbits can be achieved by introducing a difference of at least $90^\circ$ between the longitudes of the two ascending nodes of orbits with the same inclination. 

Finally, choosing the orbits of the two elements to be non-coplanar, either by changing the inclinations or the longitudes of the ascending nodes, removes directionality towards a particular orientation in the sky. This allows for denser $u-v$ plane coverage for all potential black hole targets. 

Figure~\ref{fig:2orbiters} shows the $u-v$ plane coverage for an array with two space elements that satisfy the above considerations. The semimajor axes of the two orbits are $a_1 = 14.64 \; R_{\oplus}$ and $a_2 = 12.5 \; R_{\oplus}$. Measured with respect to the equatorial plane, the orbit of the two elements have been rotated by $(i_1, \Omega_1) = (-78.39^\circ,150^\circ)$ and  $( i_2, \Omega_2) = (-90^\circ,280^\circ)$, respectively. 

We display both the $u-v$ coverage traced out for each source within one variability timescale shown in Table~\ref{tab:candidates} and the coverage that can be achieved within 6 months. There is a small level of directionality in the $u-v$ coverage for observations limited by the variability timescale. However, the full range of available azimuths are covered during the course of half a year. Figure~\ref{fig:2orbiters} demonstrates that a 2-element array with properties similar to those in this example can fulfill all of the requirements for imaging discussed earlier. For the two primary targets of the EHT (M87 and \sgra), it is worth noting that this configuration provides nearly continuous coverage in baseline lengths up to $\sim 100 {\rm G}\lambda$, which will enable the reconstruction of images with a factor of 10 higher resolution than the ground-based ones. 

Naturally, such a configuration is not unique and can be further optimized for coverage density and coverage time in the $u-v$ plane while capturing a maximum amount of information in the critical baseline lengths defined in Figure~\ref{fig:required_baselines}. However, the choices are quite limited even beyond the considerations regarding the minimum and relative altitudes of the orbits, which were discussed earlier. In particular, the are certain configurations that significantly reduce the $u-v$ coverage and should be avoided.  

One such pitfall is the case of commensurate orbital periods for the two orbital elements, which correspond to semimajor axes that obey 
\begin{equation}
    \dfrac{a_1}{a_2} = \left( \dfrac{n}{m} \right)^{2/3},
    \label{eq:commensurate}
\end{equation}
where $n$ and $m$ are positive integers. These orbital separations are non trivial and yield a repetitive closed track in the u-v plane. Its associated completion time can be expressed as
\begin{equation}
    \Delta T = 2 \pi n \sqrt{\dfrac{a_2^3}{GM_\oplus}} = 2 \pi m \sqrt{\dfrac{a_1^3}{GM_\oplus}}
\end{equation}
The number of elementary shapes depends strongly on $n$ and $m$. For small values of these integers, the $u-v$ plane will only be sparsely covered independent of the duration of the observations, as we show in the examples in Figure~\ref{fig:Commensurate}. 

\begin{figure}[t]
  \centering
  \includegraphics[width=0.7\linewidth]{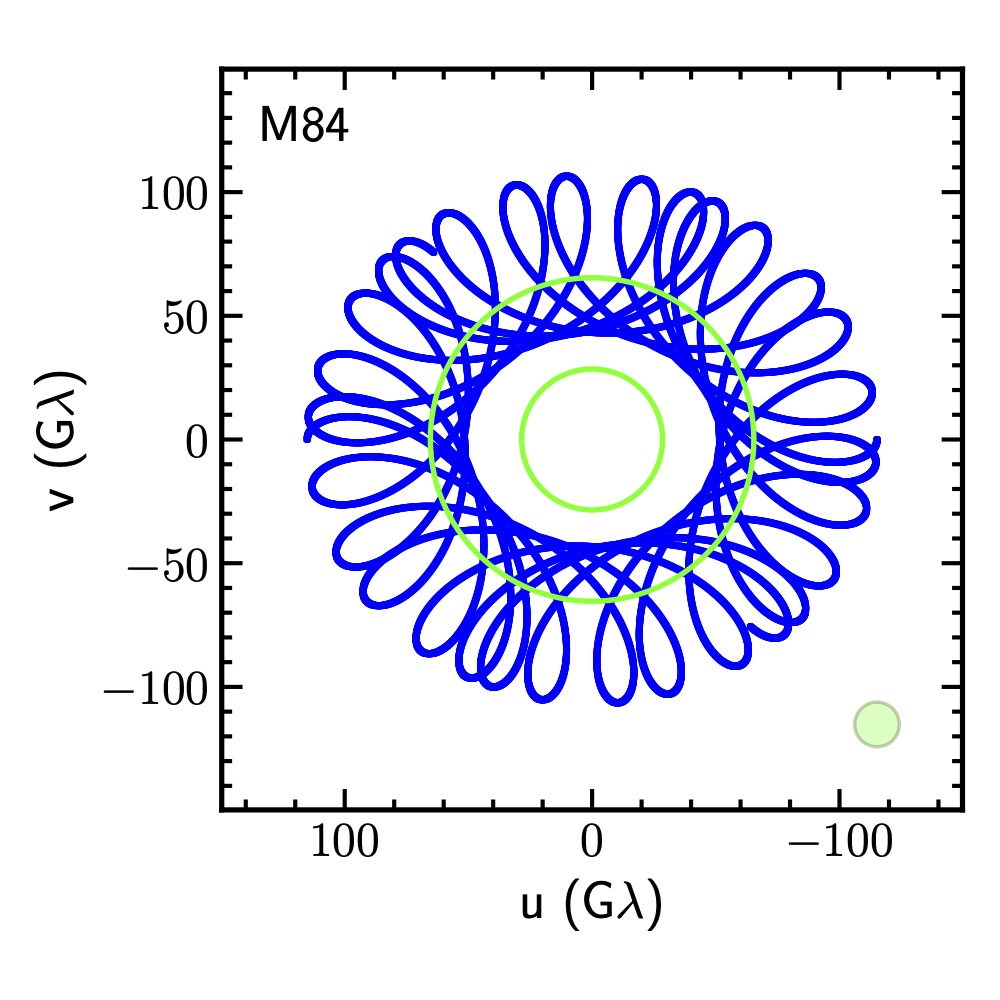}
  \caption{The effect of large separation between the orbital altitudes on the u-v plane coverage. In this example, the two elements are at semimajor axes $a_1 = 7.5 R_\oplus$ and $a_2 = 14.5 R_\oplus$, coplanar, oriented face-on to M84, and apparently counter rotating ($\Delta \Omega = 180^\circ$). This configuration fails to capture the small baseline lengths critical for determining image characteristics. }
  \label{fig:RingCoverage}
\end{figure}

A second pitfall arises from a large separation in the altitudes for the two orbital elements, which leaves large gaps in the $u-v$ coverage akin to Figure~\ref{fig:EHT_MEO}. We show such an example in Figure~\ref{fig:RingCoverage}, for which the separation between the two orbits is comparable to the orbital altitude of the first one. This choice completely misses the small baseline lengths, which are critical for determining image size and characteristics.   

\begin{figure}[t]
  \centering
  \includegraphics[width=0.49\linewidth]{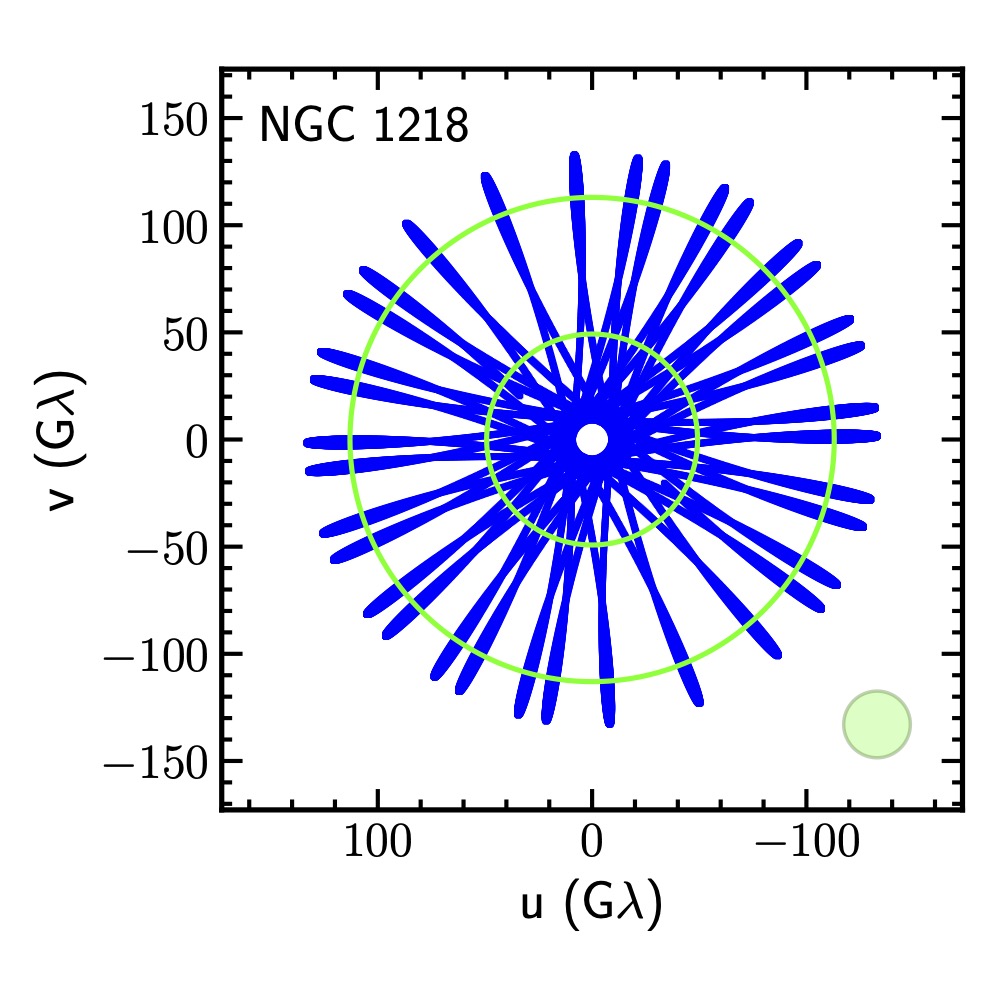}
  \includegraphics[width=0.49\linewidth]{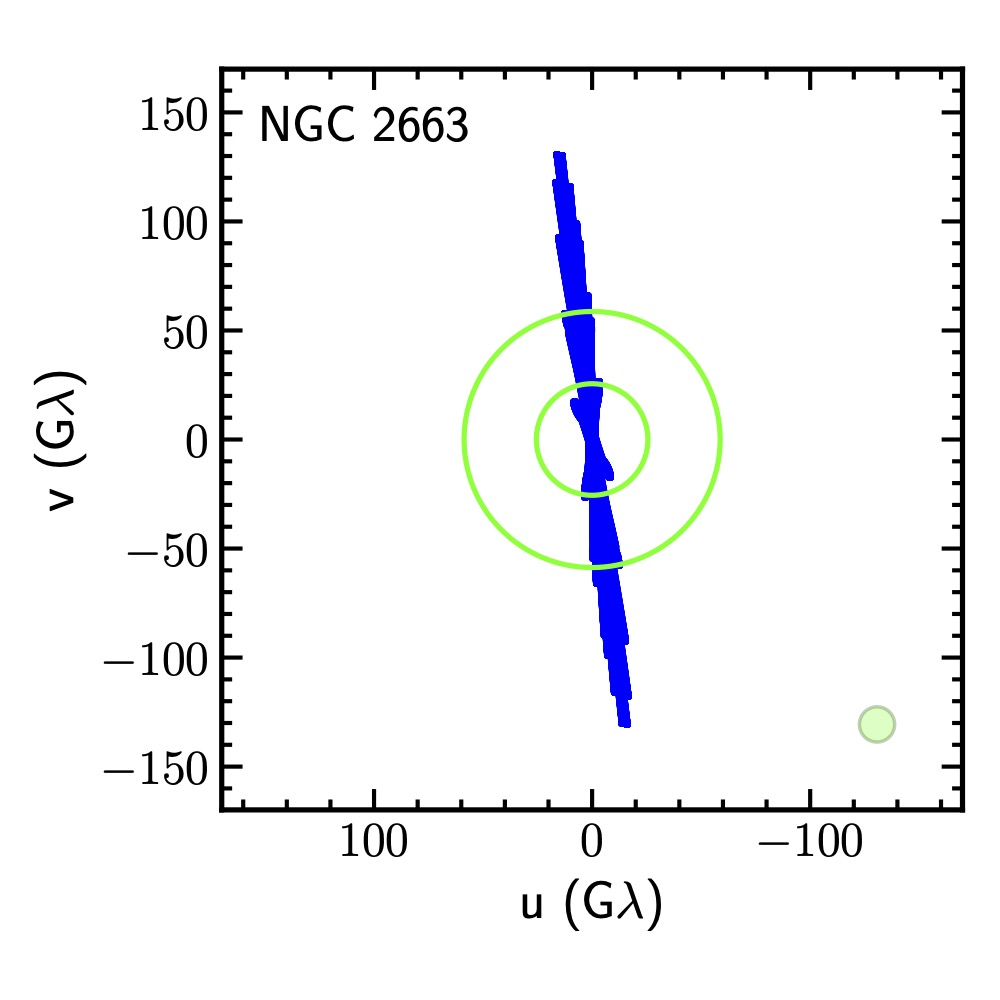}
  \caption{An example of the highly directional $u-v$ coverage resulting from a coplanar configuration. The elements are at $a_1 = 14.5 R_\oplus$ and $a_2 = 12.63 R_\oplus$ in the plane face-on to NGC 1218, with a counter rotating motion. Although this configuration optimizes coverage for NGC 1218 (left panel), it fails to achieve any azimuthal coverage for sources that are away from that plane, as in the example of NGC~2663 (right panel).}
  \label{fig:Coplanar}
\end{figure}

The central gap in such a configuration could, in principle, be reduced by aligning the projection of the orbits with the line-of-sight to the source. However, this edge-on configuration will be highly directional and cover only a very small range of azimuths in the $u-v$ plane. Co-planar orbits lead to a similar situation, becoming highly directional for sources with line-of-sights near the common orbital plane, as we show in Figure~\ref{fig:Coplanar}. 

\begin{figure}[t]
  \centering
  \includegraphics[width=0.49\linewidth]{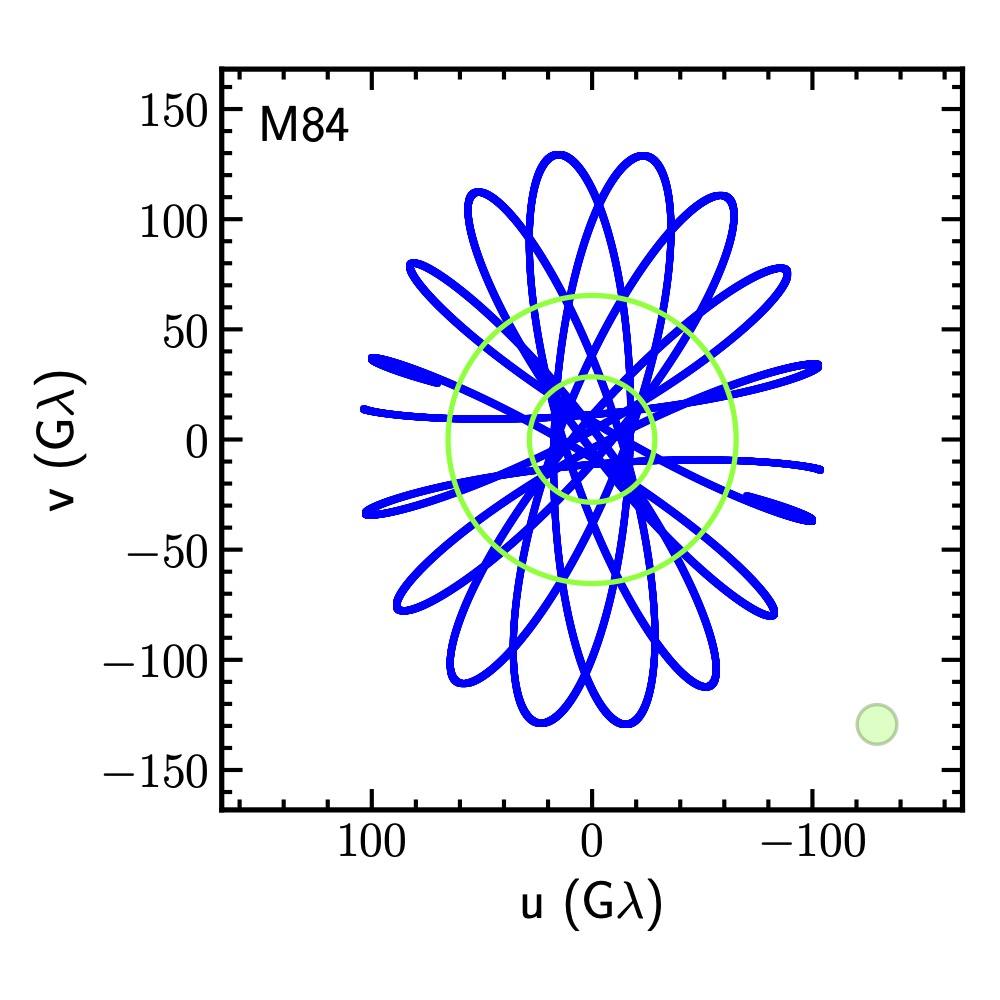}
  \includegraphics[width=0.49\linewidth]{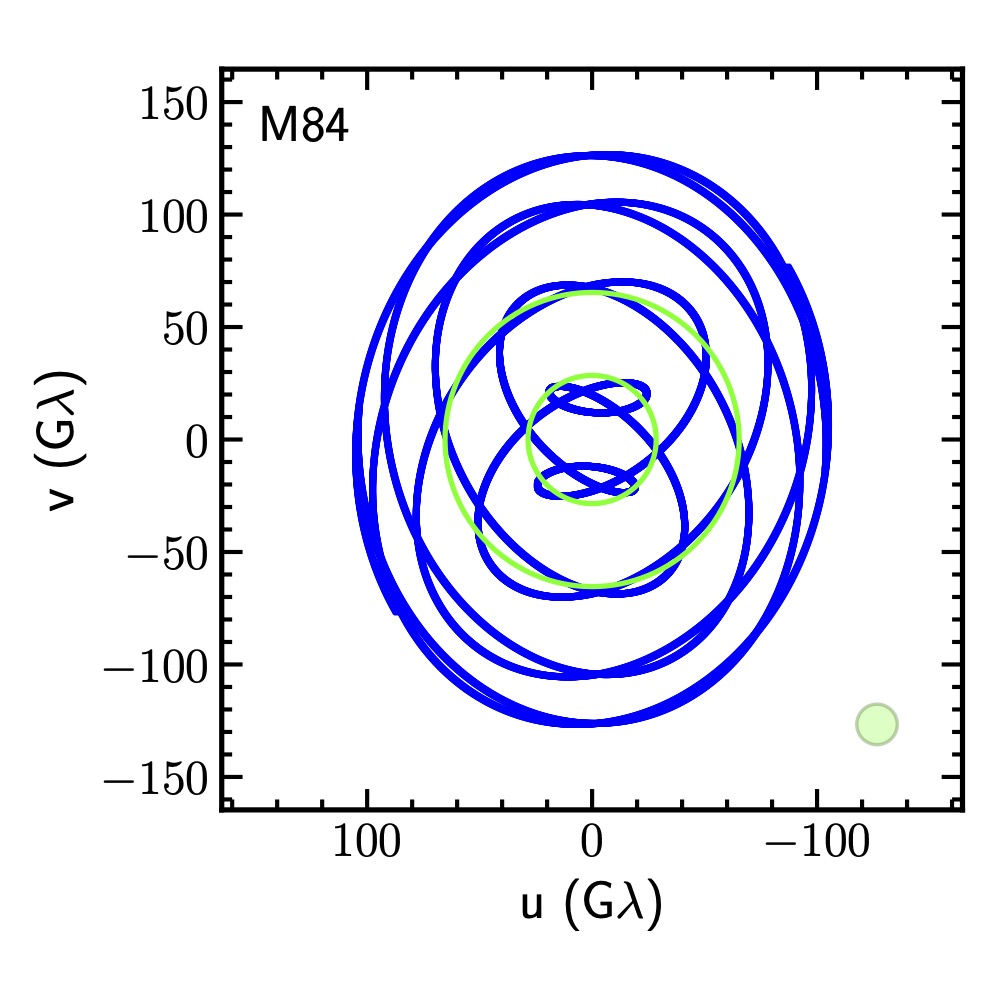}
  \caption{Comparison of the $u-v$ coverage resulting from counter-rotating elements (left) and co-rotating elements (right) or the same set up  as in Fig.~\ref{fig:2orbiters} for M84. Both the azimuthal pattern and the size of the gaps are negatively affected when the elements rotate in the same direction.}
  \label{fig:Counter}
\end{figure}

\begin{figure*}[t]
  \centering
  \includegraphics[width=0.49\linewidth]{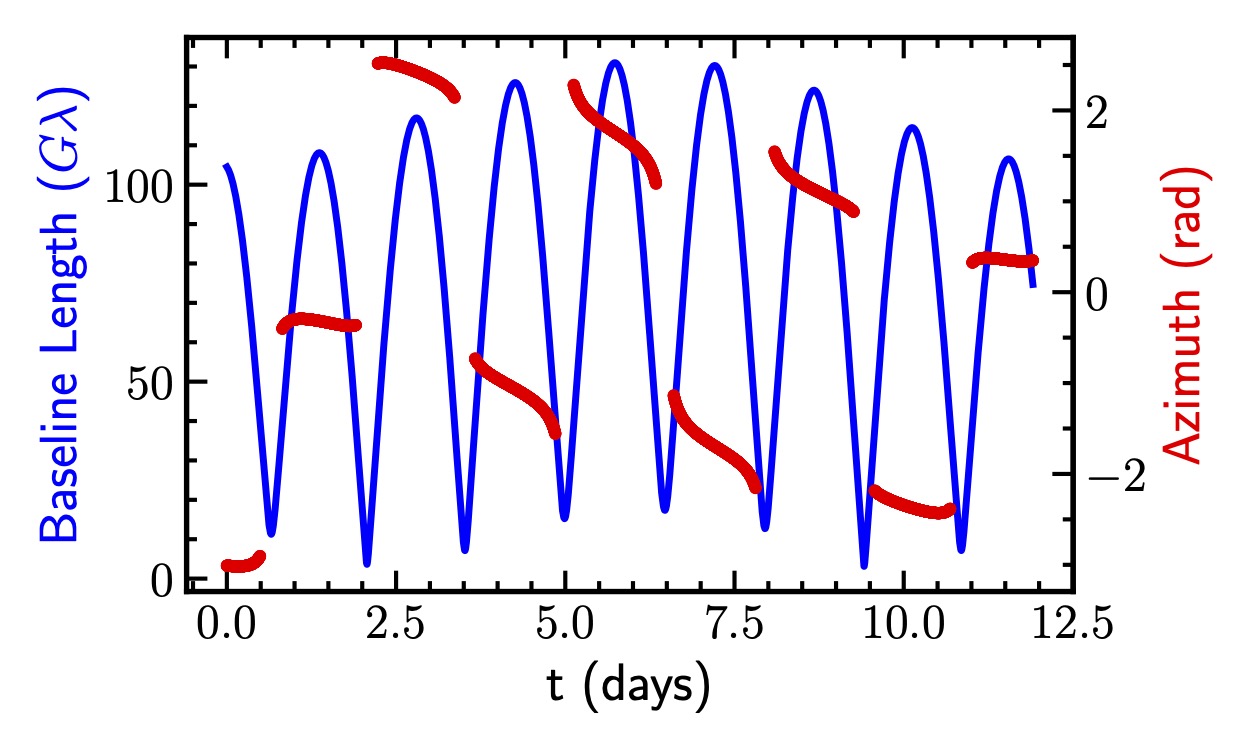}
  \includegraphics[width=0.49\linewidth]{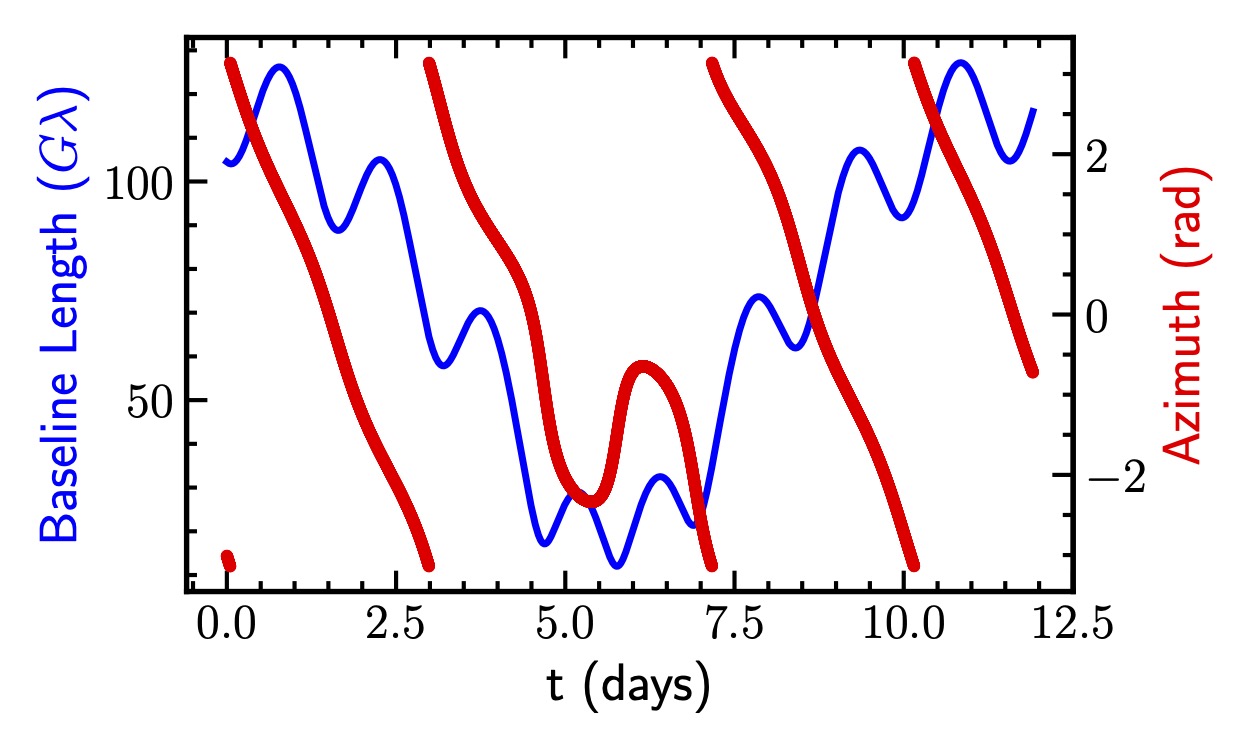}
  \caption{The radial (blue) and azimuthal (red) coverage over time for counter-rotating (left) and co-rotating elements (right) for the same configuration shown in Fig.~\ref{fig:Counter}. The counter-rotating configuration shows rapid swings in baseline length, which is advantageous for detecting potential fast changes in image structure.
  }
  \label{fig:TimeScales}
\end{figure*}

Finally, we explore the impact of the relative sense of rotation in the two orbits on imaging capabilities. In Figure~\ref{fig:Counter}, we show that elements in apparently counter-rotating orbits generate a flower-like $u-v$ coverage, whereas elements in co-rotating orbits give rise to spiral-like structures. The latter introduces larger gaps in Fourier space, which is larger than the correlation length shown in this example. Furthermore, the pattern in which the $u-v$ plane is traced out over time is significantly different between the two choices. This is shown in Figure~11, where we plot the time evolution of baseline length and azimuth for the two configurations. The counter-rotating case experiences large swings in baseline length in short periods of time (i.e., each petal in the flower-like shape), while the azimuth of the baselines slowly drifts. In contrast, for the co-rotating case, the baseline azimuth evolves rapidly, while the baseline length slowly evolves between its maximum and minimum value. Because the baseline lengths of the deep minima are most sensitive to capturing image variability, it is advantageous to have a configuration that can probe the largest range of baseline lengths in a shorter period of time. This is achievable in an apparently counter-rotating configuration.

\subsection{Configurations with Three Space Elements} 
\label{sec:TriOrbiter}

A natural approach to filling the undersampled zones of the $u-v$ plane is to add a third element to the configuration. The three-element dynamics is significantly more complex and less subject to symmetries. For example, it is easy to avoid directionality in the $u-v$ coverage of any target, independent of its position in the sky. Moreover, adding a third element leads to a denser and more uniform coverage. 

\begin{figure}[ht]
    \centering
    \includegraphics[width=0.494\linewidth]{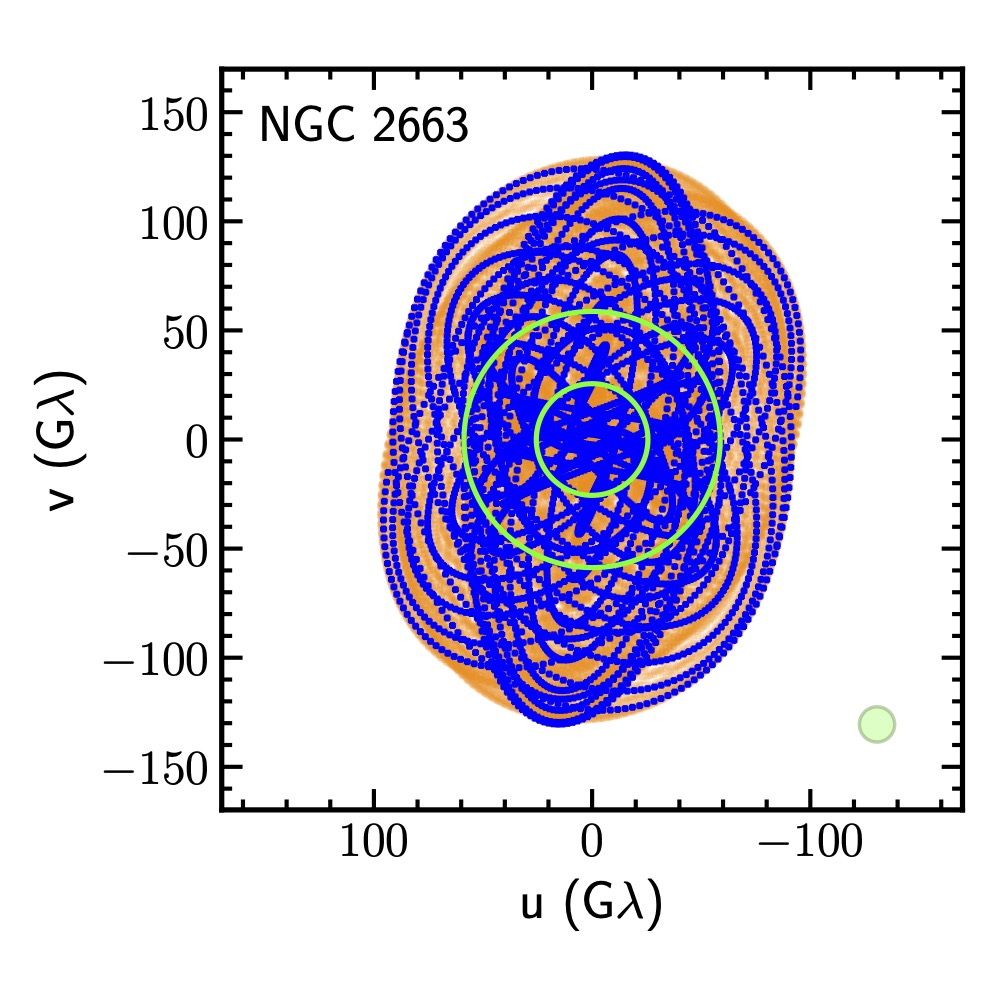}
    \includegraphics[width=0.494\linewidth]{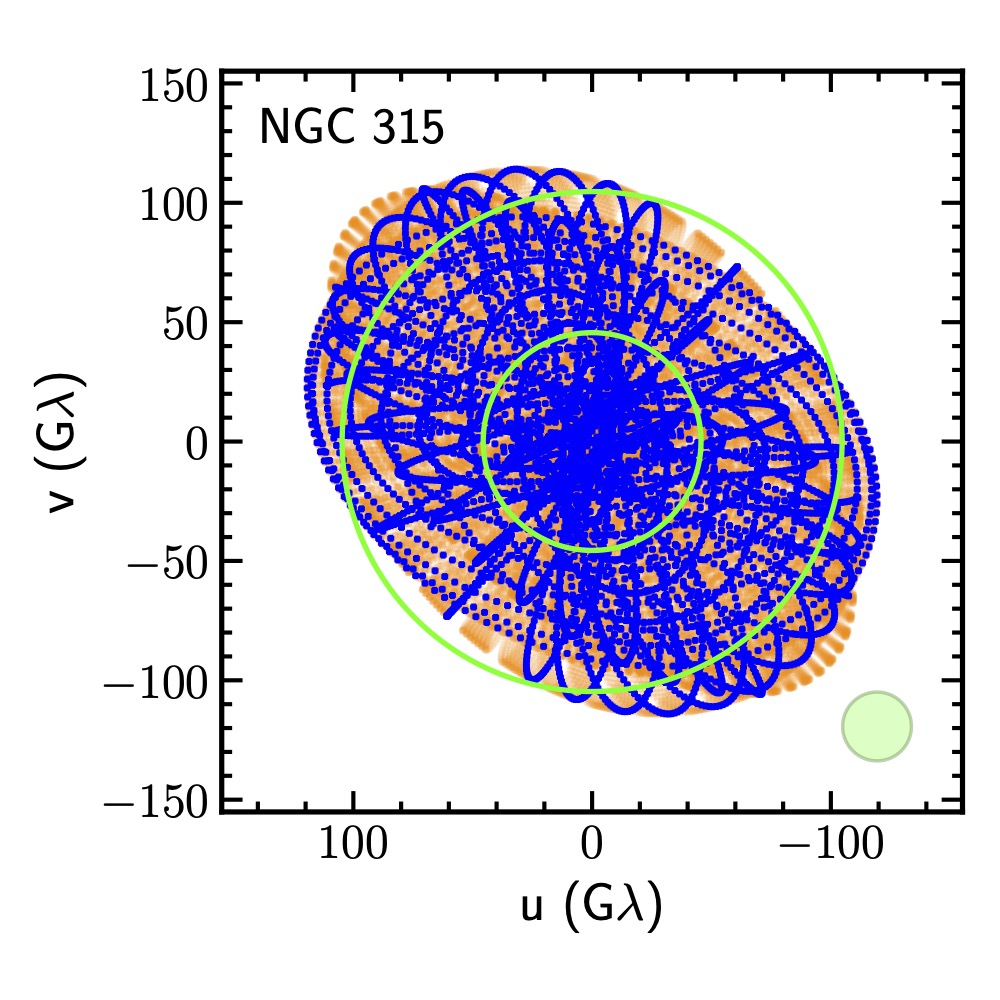}
    \caption{Examples of $u-v$ plane coverage for two targets with a three-element configuration on circular orbits at radii $a_1=14.64 R_\oplus$, $a_2 = 13.5 R_\oplus$ and $a_3 = 12.5 R_\oplus$; the first orbit has been rotated by $(i_1, \Omega_1) = (-78.39^\circ, 150^\circ)$, the second by $(i_2, \Omega_2) = (28.39^\circ, 20^\circ)$, and the third by $(i_3, \Omega_3) = (-90^\circ, 280^\circ)$. Adding a third element further reduces the directionality in coverage for some sources (cf.\ left panel with Fig.~\ref{fig:2orbiters}) and leads to even denser coverage for others.}
    \label{fig:3orbiters}
\end{figure}

To show the potential gains from adding a third array element, we start from the configuration in Figure~\ref{fig:2orbiters} and add a third element at half the inclination and longitude differences between the orbits of the first two elements. Figure~\ref{fig:3orbiters} shows the $u-v$ coverage resulting from such a configuration for two of our sources. We chose these two examples among all the targets to illustrate the largest impact a third element could introduce. As expected, the three-element configuration covers some azimuths that were incomplete for NGC~2663 and provides a denser coverage for a source like NGC~315. However, albeit denser, this $u-v$ coverage does not yield meaningfully different information about the salient features of the $u-v$ plane. In other words, a configuration with two elements already provides analogous capabilities for imaging a large number of black hole targets. For this reason, we conclude that the two-element configuration is both necessary and sufficient for black hole imaging and that adding a third element provides somewhat marginal gains in image fidelity while adding significantly to the cost of the interferometric array.

An optimization process is still necessary to find the best orbital parameters allowing for a full and dense coverage of each potential target, but due to the absence of particular orientations in most of the cases, the domain of feasibility appears way less constrained than in the case of 2-orbiters. Though such configurations provide denser and more uniform coverage, they remain highly cost inefficient provided that a 2-orbiter configuration can be found to cover at least the three high-priority regions. Such a system might be more suited to observe images with a resolution that will allow to test higher-order effects which are currently out of the scope in the field of Black Hole imaging.

\section{Conclusions}

In this paper, we explored the orbital configuration necessary for a space-based interferometric array to achieve high-fidelity imaging of supermassive black holes. We identified two key imaging requirements: sufficient baseline lengths to probe the angular sizes of target horizons and dense, continuous coverage of the $u-v$ plane in both baseline length and azimuth. These conditions are critical to reconstructing black hole images without strong priors and to capturing their essential features.

Our analysis shows that hybrid arrays combining ground-based stations with a single orbital element fail to meet these requirements due to gaps in $u-v$ coverage and limited azimuthal range; such configurations are inherently directional and fail to provide adequate imaging for a diverse set of black hole targets. Conversely, arrays with two space-based elements in high Earth orbits (HEO) overcome these challenges by introducing dense and fast $u-v$ plane coverage, particularly when the elements are in non-coplanar, apparently counter-rotating orbits. This configuration minimizes gaps in baseline lengths, ensures coverage across azimuths, and enables imaging of sources distributed throughout the sky. By excluding ground-based stations, this setup avoids the limitations imposed by atmospheric phase delays and scheduling constraints, further enhancing observational efficiency and reliability.

We demonstrated that with two space elements, the required baseline lengths can be achieved at orbital altitudes of $12-15$ Earth radii. A small difference in orbital altitudes introduces rapid drift in the relative positions of the two elements, further enhancing $u-v$ coverage over short timescales. We also identified specific failure modes in orbital choices that need to be avoided, including commensurate orbital periods, large altitude separations, and co-rotating orbits, all of which reduce coverage density and imaging capability.

Finally, we emphasize that while the example configuration presented here fulfills the requirements for imaging a wide range of black hole targets, further optimization is possible to enhance $u-v$ coverage and maximize the scientific return. Our findings provide a foundation for the design of space-based VLBI missions that significantly extend the frontiers of black hole astrophysics beyond the capabilities of the current Event Horizon Telescope.

\bibliography{MS.bib}{}
\bibliographystyle{aasjournal}

\appendix

\section{The Array of Telescopes in Orbit Simulator (ARTELOS)} \label{sec:ARTELOS}

We introduce a parametric and object oriented simulation environment for modeling space dishes, ground dishes, and their respective movement in the Earth Centered Inertial (ECI) frame over the span of a given observation time. The Array of Telescopes in Orbit Simulator (ARTELOS) calculates the $u-v$ plane coverage of any source for a given array configuration.  

We identify the position of each source in the sky by its declination ($\delta$) and right ascension ($\alpha$), such that in the Earth Centered Inertial (ECI) frame, the line of sight is defined by the direction,
\begin{equation}
    \textbf{n}_{\rm LoS} = \left(\begin{array}{c}
         \cos \delta \cos \alpha  \\
         \cos \delta \sin \alpha \\
         \sin \delta
    \end{array} \right).
    \label{eq:LoS}
\end{equation}

\subsection{Ground Based Stations}

We first implement the EHT array \citep{EHT2022_II} into ARTELOS by specifying the position of each ground dish in the ECI frame at a given time that we assign as the the origin of the time coordinate in the simulation. 

Table~\ref{tab:EHT} shows the Cartesian coordinates of the ground-based stations, with each telescope's longitude evolving at a constant rate with increasing time and latitude remaining constant. Therefore, at a given time $t$ the elevation ($\mathcal{E}$) of a source from the local horizon satisfies
\begin{equation}
    \mathcal{E}(t) = \sin^{-1} \left[\sin( \phi_0 + \omega_\oplus t) \sin \delta +
    \cos (\phi_0 + \omega_\oplus t) \cos \delta \cos (\alpha + \lambda)\right].
    \label{eq:elevation}
\end{equation}
To ensure reasonable quality of observations, we impose an elevation constraint for each telescope of 15$^\circ$.

\subsection{Space Based Stations}
\label{Appendix:SpaceStations}

We specify each telescope in space using a set of $6$ orbital parameters. Because the Earth is necessarily a focus of the orbit, two parameters define the shape of the elliptical movement in the perifocal plane, the length of the semi-major axis ($a$) and the eccentricity ($e$), whereas the true anomaly ($\theta$) defines the location of the station on its trajectory.
\begin{figure*}
    \centering
    \includegraphics[width=0.43 \linewidth]{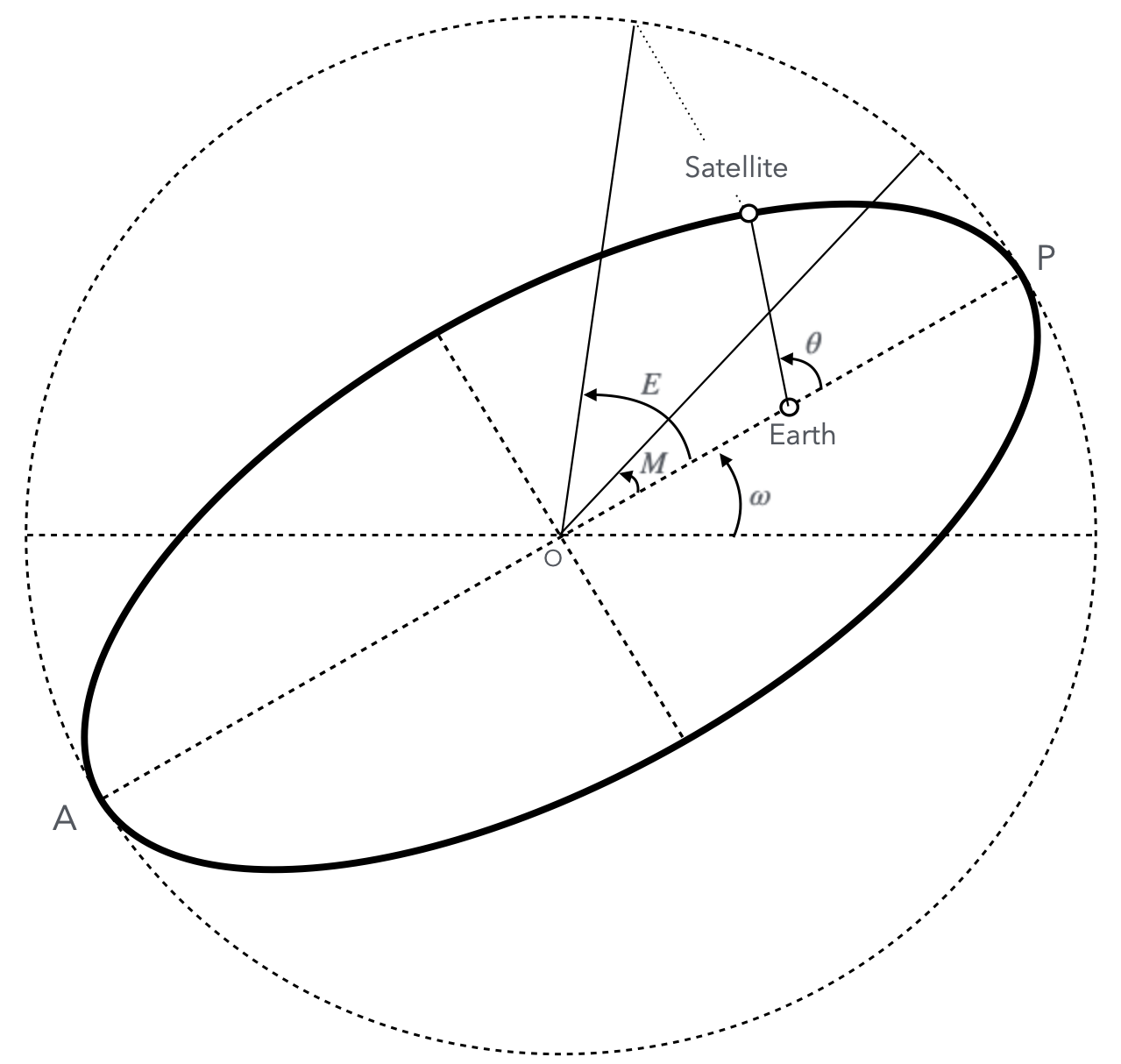}
    \includegraphics[width=0.52 \linewidth]{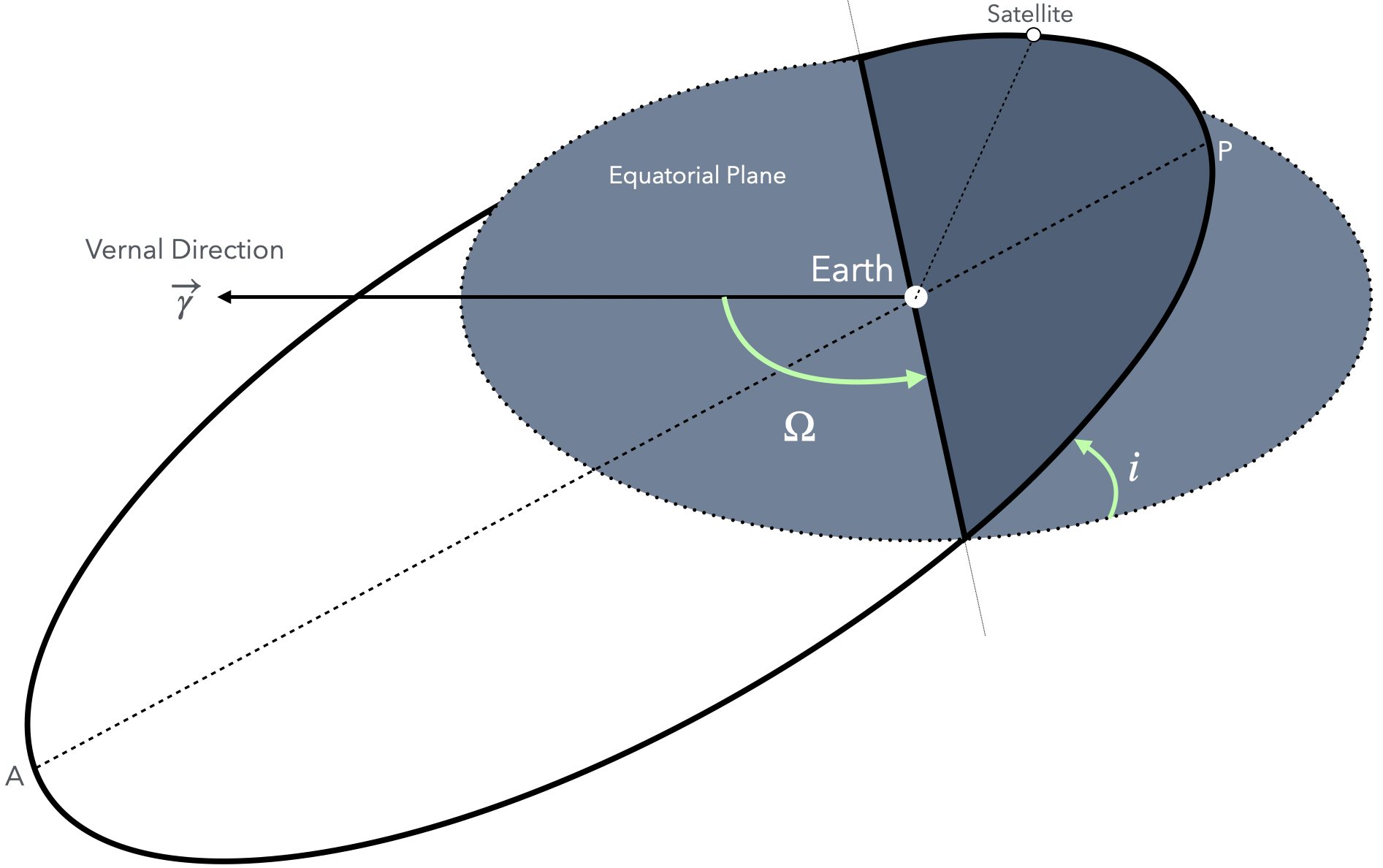}
    \caption{The orbital layout used in the calculations presented in \S\ref{Appendix:SpaceStations}. The motion in the perifocal plane (left) is characterized by the position of the periapsis (P), the apoapsis (A) and the argument of periapsis $\omega$. To compute the time evolution of the true anomaly $\theta$, we define the mean anomaly $M$ and the eccentric anomaly $E$. In the global frame (right), the inclination $i$ and the longitude of the ascending node $\Omega$ are defined relative to the equatorial plane and the vernal direction ($\overset{\rightarrow}{\gamma}$) respectively.}
    \label{Fig:OrbitalLayout}
\end{figure*}
In this plane, the orbital position vector can be expressed as
\begin{equation}
    \textbf{P} = R(a, e, \theta) \left(\begin{array}{c}  \cos \theta \\  \sin \theta \\ 0 
    \end{array}\right),
\end{equation}
where
\begin{equation}
    R(a, e, \theta) = \dfrac{a(1-e^2)}{1 + e \cos \theta}\;.
\end{equation}
In order to orient the periapsis, we rotate the whole ellipse in the perifocal plane and define the rotation angle as the argument of periapsis $\omega$. Two angles then define the orientation of the elliptical orbit in space: inclination $i$ is the angle between the perifocal plane and the equatorial plane, and the longitude of the ascending node $\Omega$ describes the angle in the equatorial plane between the ascending node and an arbitrary vernal direction. 

At any given time, these parameters can be translated into Cartesian coordinates in the ECI frame using 3D rotations of the ellipse equation in the perifocal plane. This gives
\begin{equation}
    \textbf{r} = R_Z(-\Omega) R_X(-i) R_Z(-\omega) \textbf{P},
    \label{eq:spacecoords}
\end{equation}
where $R_Z$ and and $R_X$ are the rotation matrices around the z-axis and the x-axis in the ECI frame, respectively. For more clarity, Figure \ref{Fig:OrbitalLayout} provides an overview of the concepts that we introduced.

\begin{deluxetable}{lccccc}[t!]
\tablewidth{\textheight}
\tablecolumns{4}
\tablehead{\colhead{Name}&\colhead{$X_i \, (R_\oplus)$}&\colhead{$Y_i \,(R_\oplus)$}&\colhead{$Z_i \, (R_\oplus)$}&\colhead{$\phi_0$}&\colhead{$\lambda$}}
\tablecaption{Initial Cartesian coordinates of the EHT array. \label{tab:EHT}}
\startdata
ALMA/APEX & $0.34925$ & $-0.85388$ & $-0.38953$ & $-67^\circ \, 45' \, 17.092''$ & $-22^\circ \, 53' \, 27.920''$  \\
SMT & $-0.28705$ & $-0.79335$ & $0.53804$ & $-109^\circ \, 53' \, 28.481''$ & $32^\circ \, 31' \, 37.125''$ \\
LMT & $-0.12066$ & $-0.93996$ & $-0.32387$ & $-97^\circ \, 18' \, 53.208''$ & $18^\circ \, 52' \, 4.005''$ \\
SMA/JCMT & $-0.85772$ & $-0.39129$ & $0.33759$ & $-155^\circ \, 28' \, 39.172''$ & $19^\circ \, 42' \, 6.635''$ \\
PV & $0.79877$ & $-0.04735$ & $0.60038$ & $-3^\circ \, 23' \, 33.391''$ & $36^\circ \, 52' \, 52.778''$ \\
SPT & $\sim 0$ & $\sim 0$ & $\sim -1$ & $-45^\circ \, 15' \, 0.300''$ & $-89^\circ \, 59' \, 22.900''$ \\
GLT & $0.08500$ & $-0.21786$ & $0.97017$ & $-68^\circ \, 41' \, 8.824''$ & $76^\circ \, 26' \, 51.906''$ \\
NOEMA & $0.71009$ & $0.07346$ & $0.70009$ & $5^\circ \, 54' \, 24.005''$ & $44^\circ \, 26' \, 28.868''$ \\
KP & $-0.31329$ & $-0.79067$ & $0.52693$ & $-111^\circ \, 36' \, 53.475''$ & $31^\circ \, 46' \, 50.972''$
\enddata
\end{deluxetable}

Among these orbital parameters, only the true anomaly, $\theta$, and the longitude of the ascending node, $\Omega$, evolve with time. Following standard approach, we first define the mean anomaly
\begin{equation}
    M \equiv M_0 + 2 \pi \displaystyle{\frac{t - t_0}{P_{\rm orb}}}\;.
    \label{eq:MeanAnomaly}
\end{equation}
Here, the orbital period $P_{\rm orb}$ is given by Kepler's third law
\begin{equation}
    P_{\rm orb} = \sqrt{\displaystyle{\frac{4 \pi^2 a^3}{GM_\oplus}}}\;,
\end{equation}
$M_0$ is an offset of the mean anomaly that depends on the eccentricity and the initial position of the orbiter at $t_0$, i.e.,
\begin{equation}
    M_0 \equiv \arg \left[(e + \cos \theta_0) + \boldsymbol{i}\sqrt{1-e^2} \sin \theta_0   \right] - e \displaystyle{\frac{\sqrt{1-e^2} \sin \theta_0}{1 + e\cos \theta_0}},
\end{equation}
and $\theta_0$ is the true anomaly at $t_0$. 

We then define an eccentric anomaly $E$ such that
\begin{equation}
    E - e \sin E = M\;.
    \label{eq:eccentric}
\end{equation}
Because this equation does not have a known analytical solution, we use a Newton-Raphson procedure to solve it, which converges in less than $10$ steps with a tolerance of $10^{-8}$. We use the value of $E$ obtained from equation~(\ref{eq:eccentric}) to calculate the true anomaly at time $t$,
\begin{equation}
    \theta = 2 \tan^{-1}\left[\sqrt{\displaystyle{\frac{1+e}{1-e}}} \tan \left(\displaystyle{\frac{E}{2}} \right) \right]\;.
    \label{eq:TrueAnomaly}
\end{equation}

We also incorporate the effects of nodal precession on the orbits. We write the rate of change of the longitude of the ascending node as
\begin{equation}
    \omega_p = -\displaystyle{\frac{3 \pi R_\oplus^2}{T_{orb} \left[a (1-e^2)\right]^2}} J_2 \cos i,
\end{equation}
where $J_2$ is the Earth's second dynamic factor that depends on its oblateness $\varepsilon_\oplus$ through,
\begin{equation}
    J_2 = \displaystyle{\frac{2 \varepsilon_\oplus}{3} - \frac{R_\oplus^3 \omega_\oplus}{3 G M_\oplus}}.
\end{equation}
The tides induced by the presence of the Moon make the oblateness variable and expensive to compute. The use of a Geopotential model published by the International Earth Rotation and Reference Systems Service (IERS), however, provides an estimate that incorporates the Moon induced variability, yielding
\begin{equation}
    J_2 = 1.08262668 \times 10^{-3}.
\end{equation}
In this model, the precession rate, therefore, only depends on the orbits inclination, eccentricity and semi-major axis. This effect is important only for Low Earth Orbit (LEO), as the precession time exceeds a century for altitudes beyond the Geosynchronous Orbit (GEO).

Using these definitions, we write the state vector in the perifocal plane as
\begin{equation}
    \textbf{O}(t) = \dfrac{1}{R(a, e, \theta)} \left(\begin{array}{c}
        R^2(a,e,\theta) \cos \theta (t)   \\
        R^2(a,e, \theta) \sin \theta(t) \\
        0 \\
        - \sqrt{\mu a} \sin E(t) \\
        \sqrt{1-e^2} \cos E(t) \\
        0
    \end{array} \right),
\end{equation}
where $\mu = GM_\oplus$. To obtain the orbital state vector in the ECI frame, we perform the same rotation as in equation~(\ref{eq:spacecoords})
\begin{equation}
    \textbf{S}(t) = R_Z(-\Omega - \omega_p t) R_X(-i) R_Z(-\omega) \textbf{O}(t)\;.
    \label{eq:StateVector}
\end{equation}

Finally, we take into account the observability of a target from space, which is only possible if the Earth does not block the line of sight. We define a ``cone of avoidance'' using the position vector of the orbiter in the ECI frame, $\textbf{r}_{orb}$, and the unit vector along the direction to the line of sight to the black hole in that same frame, $\textbf{n}_{\rm BH}$, such that the orthogonal projection of the position of the orbiter onto the line of sight is
\begin{equation}
    \textbf{r}_\perp = \left(\textbf{r}_{\rm orb} \cdot \textbf{n}_{\rm BH} \right) \textbf{n}_{\rm BH}\;.
\end{equation}
If the orbiter is occulted by the Earth, then
\begin{equation}
    \textbf{r}_\perp \cdot \textbf{n}_{\rm BH} < 0,
\end{equation}
and the observability criterion becomes
\begin{equation}
    \lVert \textbf{r}_{orb} - \textbf{r}_\perp \rVert \geq R_\oplus + \displaystyle{\frac{ \lVert \textbf{r}_\perp \rVert R_\oplus }{D_{\rm BH}}} \sim R_\oplus.
\end{equation}

\subsection{$u-v$ Plane Coverage}

For each array configuration, we evaluate the coverage of the $u-v$ plane to assess its imaging capabilities. We use $\textbf{r}_1$ and $\textbf{r}_2$ to denote the positions of two stations in the ECI frame observing a source of right ascension $\alpha$ and declination $\delta$. We define the displacement vector 
\begin{equation}
    \textbf{B} \equiv \textbf{r}_1 - \textbf{r}_2.
\end{equation}
To obtain the corresponding position in the $u-v$ plane, we project $\textbf{B}$ along the $\hat{\textbf{u}}$ and $\hat{\textbf{v}}$ directions defined by
\begin{equation}
    \begin{aligned}
        \hat{\textbf{u}} &= R_Z\left(\alpha - \frac{\pi}{2}\right) \hat{\textbf{X}} \\
        \hat{\textbf{v}} &= R_{\hat{\textbf{u}}}(\delta) \hat{\textbf{Z}}, 
    \end{aligned}
\end{equation}
which finally yields the following expressions
\begin{equation}
\begin{aligned}
         \lambda_{obs} u =& \textbf{B} \cdot \hat{\textbf{u}} \\ =& \,(X_1 - X_2) \sin \alpha + (Y_2 - Y_1) \cos \alpha  \\
         \lambda_{obs} v =& \textbf{B} \cdot \hat{\textbf{v}} \\ =& (X_2 - X_1) \sin \delta \cos \alpha + (Y_2 - Y_1) \sin \delta \sin \alpha + (Z_1 - Z_2) \cos \delta.
\end{aligned}
\end{equation}
Here $\lambda_{\rm obs}$ is the wavelength of observation such that $u$ and $v$ are expressed in units of $\lambda_{\rm obs}$.

\end{document}